
\input amstex
\documentstyle{amsppt}
\NoRunningHeads
\TagsOnRight
\magnification=1200
\document
$$\qquad\qquad\qquad\qquad\qquad\qquad\qquad\qquad\qquad\qquad\qquad\qquad
\text{LMU-TPW 93-16}$$
\bigskip
\bigskip
\bigskip
$$\gather\text{{\bf UNIVERSAL BUNDLE FOR GRAVITY, LOCAL INDEX THEOREM,}}\\
\text{{\bf AND COVARIANT GRAVITATIONAL ANOMALIES}}\endgather$$
$$\gather\text{Gerald Kelnhofer	\footnotemark"*"}\\ \text{Sektion Physik
der Ludwig Maximilians Universit\"at}\\ \text{Theresienstr. 37, 8000
M\"unchen 2}\\\text{Germany}\endgather$$
\bigskip
\bigskip
\bigskip
\bigskip
\head Abstract\endhead
\footnotetext"*)"{Erwin Schr\"odinger fellow, supported by "Fonds zur
F\"orderung der wissenschaftlichen Forschung in \"Osterreich", project
number: J0701-PHY}%
Consistent and covariant Lorentz and diffeomorphism anomalies are
investigated in terms of the geometry of the universal bundle for gravity.
This bundle is explicitly constructed and its geometrical structure will be
studied. By means of the local index theorem for families of Bismut and Freed
the consistent gravitational anomalies are calculated. Covariant
gravitational anomalies are shown to be related with secondary
characteristic classes
of the universal bundle and a new set of descent equations which also
contains the
covariant Schwinger terms is derived. The relation between consistent and
covariant anomalies is studied. Finally a geometrical realization of the
gravitational BRS, anti-BRS transformations is presented which enables the
formulation of a kind of covariance condition for covariant gravitational
anomalies.
\par\newpage
{\bf I. Introduction}
\bigskip
It has been shown long $\text{ago}^{{}1,2}$ that certain chiral fermions
interacting
with external gravitational fields in $4k+2$ dimensions may suffer from
anomalies which spoil general covariance and local Lorentz symmetry. These
gravitational anomalies appear as a non-conservation of the energy momentum
tensor and cause its asymmetry. \par
Like in the Yang-Mills case one can distinguish between a consistent and
covariant form of the gravitational anomaly. Actually, the
methods used in Ref. 1 give the covariant diffeomorphism anomaly. On the
other
side, pure covariant Lorentz anomalies have been calculated in Ref. 3 using
path-integral methods. Bardeen and $\text{Zumino}^{{}4}$  investigated the
relation
between the consistent and covariant diffeomorphism anomaly. Moreover they
showed that consistent diffeomorphism anomalies can be shifted to consistent
Lorentz anomalies by means of a non-polynomial action.
\par
Another manifestation of these anomalies is due to the occurance of
Schwinger terms in the equal time commutator of the energy momentum tensor.
The consistent Schwinger terms have been determined in two dimensions and
their relation with the gravitational anomaly was $\text{investigated.}^{{}
5,6}$\par
Recently, an algebraic approach to determine covariant diffeomorphism
anomalies has been $\text{presented.}^{{}7,8}$ This method is based on the
gauged (anti) BRST symmetry
and leads to descent equations for covariant diffeomorphism anomalies and
Schwinger terms. Furthermore a kind of consistency condition for these
anomalous terms has been formulated in Ref. 8. \par
The aim of this paper is twofold: In the first place we want to clarify the
applicability of the local index theorem of Bismut and $\text{Freed}^{{}
9-11}$ to
the determination of consistent gravitational anomalies. In the second place
we intend
to investigate the geometrical structure of covariant anomalies.\par
Although the geometrical framework and the use of the cohomological
Atiyah-Singer index $\text{theorem}^{{}12}$ to the study of the consistent
case has been outlined by various $\text{authors,}^{{}13-16}$
a throughout construction has never been worked out explicitly. \par
Since local $\text{cohomology}^{{}16}$ seems to be the appropriate
cohomology theory to detect
anomalies, the use of the Atiyah-Singer index theorem needs further
clarification. In the Yang-Mills case Atiyah and $\text{Singer}^{{}13}$
introduced a certain universal bundle with connection in order to compute
the characteristic classes of the index bundle. Its first Chern class was
interpreted as an obstruction to trivializing the associated determinant
line bundle. It was argued that a similar situation arises in the
gravitational case.\par
On the other
hand the local families index theorem of Bismut and  $\text{Freed}^{{}10,
11}$
provides a method to calculate the curvature of the determinant line bundle
associated with the family of Dirac operators. The relevance of this
differential form version of the index theorem in String theory has been
stressed by $\text{Freed.}^{{}17}$ He also sketched briefly the
gravitational case
but he skipped an explicit calculation.\par
It
is one aim of this paper to fill this gap in order to study consistent and
covariant gravitational anomalies form a unified geometrical viewpoint. We
shall give an explicit construction of the universal bundle for gravity and
define a universal connection in the sense of Ref. 13 on it. It is then shown
that this bundle with connection provides the geometrical data in order to
apply the local index theorem. Finally we calculate the consistent
gravitational anomalies.\par
In view of possible applications in topological gravity a
detailed investigation of the geometrical structure of the universal bundle
seems to be interesting in itself. A step in this direction has been already
made in two dimensions in Ref. 18.\par
Our other theme is to investigate the geometrical structure of combined
covariant diffeomorphism- and Lorentz anomalies of arbitrary ghost degree.
Here we shall present a construction extending
the formalism which we have introduced in the case of Yang-Mills
$\text{theory.}^{{}19}$
This method is based on the geometry of the universal bundle for
gravity and unlike other $\text{formalisms}^{{}7,14}$ it does not require
the use of
certain homotopy operators. Moreover our method is appropriate to deal also
with non parallelizable manifolds. A further
advantage of our formalism is that it povides explicit formulae for
consistent and covariant anomalies and their corresponding counter terms.\par
We shall show that covariant
gravitational anomalies are related with secondary characteristic classes
of a certain pullback bundle of the universal gravity bundle. Furthermore
a new set of descent equations for the covariant gravitational anomalies of
arbitrary degree in diffeomorphism- and Lorentz ghost can be derived. \par
As a by-product of our study we shall see that the covariant
diffeomorpism anomaly can be viewed as an equivariant momentum map on the
space of all metrics. In the case of covariant Yang-Mills anomalies a similar
result has already been obtained in Ref. 20.\par
Finally we find a geometrical realization of the gravitational BRS, anti-BRS
$\text{multiplet.}^{{}21}$ The covariant descent equations can be formulated
in this extended framework and it is shown
that the covariant gravitational anomalies satisfy a kind of covariance
condition. For pure diffeomorphism anomalies we recover recent results of
Ref. 8.
\par
The paper is organized as follows: In Sec. II the universal bundle for
gravity is constructed. Its geometry will be discussed in detail and a
natural connection in the sense of Ref. 13 will be constructed. In Sec.
III, we present the geometrical framework to dicuss both diffeomorphisms and
local Lorentz transformations and derive the gravitational BRS multiplet.
The family of Dirac operators is constructed and by means of the local
families index theorem we calculate the curvature of the corresponding
determinant line bundle. Then the consistent gravitational anomalies are
determined. Sec. IV is devoted to the determination of covariant
gravitational anomalies. A new set of descent equations is derived and
finally the relation between consistent and covariant anomalies is
investigated. A geometrical realization of the gravitational BRS, anti-BRS
multiplet is formulated in Sec. V. It is shown that the covariant anomalies
fulfill a certain covariance condition.
\bigskip\bigskip
{\bf II. The universal bundle for gravity}
\bigskip
Let $M$ be a compact, connected, orientable, $n$-dimensional spin manifold
without boundary. Let $LM^{+}$ denote the principal bundle of oriented
linear frames over $M$ with structure group $GL^+(n)$ and projection
$\pi _{LM^+}$. A frame in $x\in M$ is an isomorphism $u\colon
\Bbb R^n\rightarrow T_xM$. The right action by
$A\in GL^+(n)$ is denoted by $r_A(u):=u\cdot A$ and the Lie algebras of
$GL^+(n)$ and $SO(n)$ will be denoted by $gl(n)$ and $so(n)$, respectively.
\par
The group of orientation preserving diffeomorphisms of $M$ will be denoted
by $\frak D$. Let us
consider the group of bundle automorphisms $Aut(LM^+)$ of $LM^+$. The
subgroup of $Aut(LM^+)$ which induces the identity transformation on $M$ is
denoted by $Aut_0(LM^+)$. There is the following exact sequence of groups
$$1\rightarrow Aut_0(LM^+)\rightarrow Aut(LM^+)\rightarrow\frak D\rightarrow
1.\tag2.1$$
Define $l\colon\frak D\rightarrow Aut(LM^+)$ by $l(\phi)u:=T_{\pi (u)}\phi
\circ u$ then $Aut(LM^+)$ can be regarded as a semidirect product of
$\frak D$ and $Aut_0(LM^+)$, denoted by $\frak D\ltimes Aut_0(LM^+)$,
relative
to the homomorphism $j\colon\frak D\rightarrow Aut(Aut_0(LM^+))$ which is
given by
$$j(\phi )(F):=l(\phi )\circ F\circ l(\phi ^{-1}),\qquad\phi\in\frak D,\ F\in
Aut_0(LM^+).\tag2.2$$
Correspondingly, any vector field $\zeta\in\frak X(M)$ can be lifted to a
unique $GL^+(n)$ invariant
vector field $Y^{\zeta}\in\frak X(LM^+)$ given by $Y_u^{\zeta}:=\frac{d}{dt}
\vert _{t=0}
l(Fl_t^{\zeta})u$, where $Fl_t^{\zeta}$ is the flow generated by $\zeta\in
\frak X(M)$.
Notice that $Y^{\zeta}$ can be considered as infinitesimal generator of the
left action $(\phi ,u)\mapsto l(\phi )u$ of $\frak D$ on $LM^+$.
Therefore one has $[Y^{\zeta _1},Y^{\zeta _2}]=Y^{[\zeta _1,\zeta _2]}$ for
all $\zeta _1,\zeta _2\in\frak X(M)$.\par
Let
$S^2T^{\ast}M$ denote the space of symmetric covariant tensors then the space
of metrics $\frak M$ on $M$ is defined to be the set of all sections $\Gamma
(S^2T^{\ast}M)$ of $S^2T^{\ast}M$ which induce a positive inner product in
any tangent space $T_xM$. So $\frak M$ is an open cone in $\Gamma
(S^2T^{\ast}M)$ and therefore it is contractible. See Ref. 22 for a detailed
analysis of the structure of $\frak M$.\par
In this paper we shall often regard tensor fields as equivariant functions
on $LM^+$. Thereby we follow the calculus developed in Ref. 23. Let $T^{p,q}$
be the tensor algebra of $\Bbb R^n$. A representation $GL^+(n)\rightarrow
GL(T^{p,q})$ is given by
$$(A\cdot f)(a_1,\ldots ,a_p,v_1,\ldots v_q):=f(a_1\circ A,\ldots ,a_p
\circ A,A^{-1}(v_1),\ldots ,A^{-1}(v_q)),\tag2.3$$
where $f\in T^{p,q},\ a_i\in (\Bbb R^n)^{\ast},\ v_i\in\Bbb R^n$ and
$(\Bbb R^n)^{\ast}$ is the dual space of $\Bbb R^n$. Then the space
$$C(LM^+,T^{p,q}):=\lbrace f\colon LM^+\rightarrow T^{p,q}\vert
f(u\cdot A)=(A^{-1}\cdot f)(u)\rbrace\tag2.4$$
is isomorphic with the space $\frak T^{p,q}=\Gamma (\otimes ^pTM
\otimes\otimes ^qT^{\ast}M)$ of tensor fields of type $(p,q)$ on $M$. Let
$t\in\frak T^{p,q}$ then we denote the corresponding equivariant function by
$\hat t$. Conversely, an equivariant function $\hat t$ determines $t$ by
the formula
$$t_{\pi (u)}(\alpha _1,\ldots ,\alpha _p,\zeta _1,\ldots ,\zeta _q):=
\hat t(u)(\alpha _1\circ u,\ldots ,\alpha _p\circ u,u^{-1}(\zeta _1),\ldots ,
u^{-1}(\zeta _q)),\tag2.5$$
where $\zeta _i\in T_{\pi (u)}M$ and $\alpha _i\in T_{\pi (u)}^{\ast}M$.
Let $(e_i)$ be the
standard basis of $\Bbb R^n$ and $(e^i)$ its dual basis, then the
components of $f\in C(LM^+,T^{p,q})$ are the equivariant functions
$$f_{j_1\ldots j_q}^{i_1\ldots i_p}(u):=f(u)(e^{i_1},\ldots ,e^{i_p},
e_{j_1},\ldots ,e_{j_q}).\tag2.6$$
Let $\Gamma (g)\in\Omega ^1(LM^+,gl(n))$ be the Levi-Civita connection
in $LM^+@>>> M$ associated with the metric $g$. Let us recall that
$\Gamma (g)$ is uniquely determined by the equations
$$d_{\Gamma (g)}\ \hat g=0,\qquad d_{\Gamma (g)}\ \varphi =0,
\tag2.7$$
where $\varphi\in\Omega ^1(LM^+,\Bbb R^n)$ is the soldering form, i.e.
$\varphi _u(X_u)=u^{-1}(T_u\pi _{LM^+}(X_u))$. Here
$d_{\Gamma (g)}$ denotes the covariant exterior derivative on
$LM^+$ with respect to $\Gamma (g)$. For any $v\in\Bbb R^n$ there exists
a unique horizontal (relative to $\Gamma (g)$) vector field $\bar v\in
\frak X(LM^+)$ such that $\varphi _u((\bar v)_u)=v$ [31]. The components
of $d_{\Gamma (g)}\hat t=d_{LM^+}\hat t+\Gamma (g)\cdot\hat t$, where
$t\in\frak T^{p,q}$ are given by
$$(d_{\Gamma (g)}\hat t)_{j_1\ldots j_qk}^{i_1\ldots i_p}\equiv
\hat t_{j_1
\ldots j_q;k}^{i_1\ldots i_p}=i_{\bar e_k}\ (d_{LM^+}\ \hat t_{j_1\ldots j_q}
^{i_1\ldots i_p}),\tag2.8$$
where $d_{LM^+}$ is the exterior derivative on $LM^+$. For $\phi\in\frak D$,
(2.7) implies
$$\Gamma (\phi ^{\ast}g)=l(\phi )^{\ast}\Gamma (g).\tag2.9$$
Let $\Omega _{eq,h}(LM^+,T^{p,q})$ denote the algebra of
equivariant (relative to (2.3)) and horizontal (relative to $\Gamma (g)$)
differential forms on $LM^+$ then there is a natural $\text{isomorphism}^{{}
23}$
$$\gather\lambda\colon\Omega _{eq,h}^1(LM^+,T^{p,q})\rightarrow
C(LM^+,T^{p,q+1})\\ \lambda (f)(u)(a_1,\ldots ,a_p,v_1,\ldots ,v_{q+1}):=
f_u((\bar v_{q+1})_u)(a_1,\ldots ,a_p,v_1,\ldots, v_q).\tag2.10\endgather$$
Since $SO(n)$ is the maximal compact subgroup of $GL(n)^+$, $LM^+$ admits a
reduction to a principal $SO(n)$ bundle. Each reduction corresponds uniquely
to a global section in the associated fibre bundle $LM^+\times _{GL(n)^+}
(GL(n)^+/SO(n))$ over $M$ and is in a one to one correspondence with a
Riemann metric on $M.^{{}24}$ But $LM^+\times _{GL(n)^+}(GL(n)^+/SO(n))$ is
isomorphic to the fibre bundle $LM^+/SO(n)$ over $M$ where frames differing
by an orthonormal transformation are identified. Thus we have the
identification $\frak M\cong\Gamma (LM^+/SO(n))$. Note that $LM^+/SO(n)$
itself is the base of the principal $SO(n)$ bundle $LM^+@>\tilde\pi >>
LM^+/SO(n)$. In the following we shall write $\widetilde{LM}^+:=
LM^+/SO(n)$.\par
So every $g\in\frak M$ induces the pullback bundle $SO_g(M):=
g^{\ast}LM^+$ over $M$ which covers all frames of $LM^+$ which are
orthonormal with respect to $g$. We introduce the evaluation map
$ev\colon\frak M\times M\rightarrow\widetilde{LM}^+$, $ev(g,x)=[u,
\hat g(u)]\in
LM^+\times _{GL^+(n)}GL^+(n)/SO(n)$ with $\pi (u)=x$, where we have
identified the homogenous space $GL^+(n)/SO(n)$ with the space of
symmetric transformations of $\Bbb R^n$. Hence $ev$ induces the
pullback bundle
$$\CD \frak F:=ev^{\ast}LM^+ @>>> LM^+\\ @VVV @V\tilde\pi VV \\ \frak M
\times M @>ev >> \widetilde{LM}^+\endCD\tag2.11$$
By definition $\frak F=\lbrace (g,x,u)\in\frak M\times M\times LM^+\vert\
\text{u is orthonormal with respect to $g_x$}\rbrace$. Obviously, $\frak F$
restricts to $SO_g(M)$ on $\lbrace g\rbrace\times M$, i.e. $\frak F\vert _
{\lbrace g\rbrace\times M}\cong SO_g(M)$.\par
If $\hat g\in C(LM^+,T^{0,2})$ corresponds to $g\in\frak M$ then we admit
an equivalent characterization of $ev^{\ast}LM^+$ by
$$\frak F=\lbrace (g,u)\in\frak M\times LM^+\vert\ \hat g(u)(v_1,v_2)
=(v_1,v_2)_{\Bbb R^n}\ \forall v_1, v_2\in \Bbb R^n\rbrace ,\tag2.12$$
where $(,)_{\Bbb R^n}$ denotes the standard scalar product on $\Bbb R^n$.
The principal right action is given by $r_O(\hat g,u):=(\hat g,u\cdot O)$
with $O\in SO(n)$. The tangent bundle of $\frak F$ is given by
$$T\frak F=\bigcup _{(g,u)\in\frak F}\ \lbrace (s_g,X_u)\in T_{(g,u)}
(\frak M\times LM^+)\vert\ \hat s_g(u)=-d_{LM^+}\ \hat g(X_u)\rbrace .
\tag2.13$$
The bundle $\frak F$ can also be obtained in the following way:
Consider the principal $SO(n)$ bundle $\frak M\times LM^+@>
\tilde\pi >>\frak M\times\widetilde{LM}^+$ with projection $\tilde
\pi (g,u):=(g,[u])$. Here $[u]$ is the equivalence class in
$\widetilde{LM}^+$. Let $\chi\colon\frak M
\times M\rightarrow\frak M\times \widetilde{LM}^+$ be defined by
$\chi (g,x):=(g,ev(g,x))$ then  $\chi ^{\ast}(\frak M\times LM^+)\cong
\frak F$.\par


The vector bundle $\frak M\times TM\rightarrow\frak M\times M$, which is
associated with the principal bundle $\frak M\times LM^+$, admits a natural
metric, given by
$$\nu _{(g,x)}((0_g,\zeta _x^1),(0_g,\zeta _x^2)):=g_x(\zeta _x^1,
\zeta _x^2),\quad\zeta ^1,\zeta ^2\in TM.\tag2.14$$
Equivalently, $\nu$ can be described by the $GL^+(n)$ equivariant map $\hat
\nu\colon\frak M\times LM^+\rightarrow T^{0,2})$, defined by
$\hat\nu (g,u):=\hat g(u)$. Hence $\frak F$ can be regarded
as the bundle of orthonormal frames of $\frak M\times TM$. Let $i\colon
\frak F\hookrightarrow\frak M\times LM^+$ denote the corresponding embedding.
\par
In the following we restrict to the group $\frak D_0$ of orientation
preserving diffeomorphisms which leave a point and a frame fixed, namely
$$\frak D_0:=\lbrace\phi\in\frak D\vert\ \phi (x_0)=x_0,\ T_{x_0}\phi
=id_{T_{x_0}M}\rbrace .\tag2.15$$
Then the right action $R^{\frak M}(g,\phi ):=\phi ^{\ast}g$ of $\frak D_0$
is $\text{free}^{{}22}$
and thus we can consider the principal $\frak D_0$ bundle $\frak M@>\pi _
{\frak M}>>\frak M/\frak D_0$. (Notice that $\phi ^{\ast}g$ corresponds to
$l(\phi )^{\ast}\hat g$ in the equivariant description.)\par
There is a free right action of $\frak D_0$ on the two principal bundles
$\frak M\times LM^+\rightarrow\frak M\times\widetilde{LM}^+$ and
$\frak M\times LM^+\rightarrow\frak M\times M$ defined by
$$\split & w_{\phi}\colon\frak M\times LM^+\rightarrow\frak M\times LM^+,\
w_{\phi}(g,u):=
(\phi ^{\ast}g,l(\phi )^{-1}u),\\ & w_{\phi}^{\prime}\colon\frak M\times
\widetilde{LM}^+\rightarrow \frak M\times\widetilde{LM}^+,\ w_{
\phi}^{\prime}(g,[u])=(\phi ^{\ast}g,[l(\phi ^{-1})u]\\ &
\bar w_{\phi}\colon\frak M\times M\rightarrow \frak M\times M,\
\bar w_{\phi}(g,x)=(\phi ^{\ast}g,\phi ^{-1}(x)).\endsplit\tag2.16$$
Finally this action extends to a bundle action on $\frak F$. Since
$$\split & w_{\phi}^{\prime}\circ \chi (g,x)=(\phi ^{\ast}g,[l(\phi
^{-1})u,\hat g(u)])=(\phi ^{\ast}g,[l(\phi ^{-1}u,(l(\phi )^{\ast}\hat g
(l(\phi ^{-1})u)])\\ &=(\phi ^{\ast}g,ev(\phi ^{\ast}g,\phi ^{-1}(x)))=
\chi\circ\bar w_{\phi }(g,x)\endsplit\tag2.17$$
the map $\chi$ factorizes to a map $\chi ^{\prime}\colon\frak M\times _{
\frak D_0}M\rightarrow\frak M\times _{\frak D_0}\widetilde{LM}^+$.
Taking the quotient by the
various $\frak D_0$ actions we obtain the following commutative diagram:
$$\CD \frak M\times LM^+ @<\bar\chi << \frak F @>\bar p>>\frak F/\frak D_0
@>\bar\chi ^{\prime}>> \frak M\times _{\frak D_0}LM^+\\ @V\tilde\pi VV
@V\pi _{\frak F} VV @V\bar\pi _{\frak F} VV @V\tilde\pi ^{\prime}VV \\
\frak M\times \widetilde{LM}^+ @<\chi << \frak M\times M @>p>>\frak M
\times _{
\frak D_0}M @>\chi ^{\prime}>>\frak M\times _{\frak D_0}\widetilde{LM}^+\\
@VVV @VVV @VVV @VVV \\ \frak M @= \frak M @>\pi _{\frak M}>>\frak M/\frak D_0
@= \frak M/\frak D_0\endCD\tag2.18$$
Here $p$, $\bar p$ are the projections and $\bar\chi$ and $\bar\chi ^{
\prime}$ are the canonical bundle maps. Notice that $\frak M\times _{
\frak D_0}M$, $\frak M\times _{\frak D_0}\widetilde{LM}^+$ and $\frak M
\times _{\frak D_0}LM^+$ can be considered as fibre bundles associated with
$\frak M\rightarrow\frak M/\frak D_0$, where the fibres are $M$, $\widetilde
{LM}^+$ and $LM^+$, respectively. Obviously one finds the bundle
isomorphisms
$$p^{\ast}(\frak F/\frak D_0)\cong\frak F,\qquad\chi ^{\prime\ \ast}(\frak M
\times _{\frak D_0}LM^+)\cong \frak F/\frak D_0.\tag2.19$$
Since $\frak D_0$ acts as an isometry of $\nu$, this metric factorizes to a
metric $\bar\nu$ on $\frak M\times _{\frak D_0}TM\rightarrow\frak M\times _{
\frak D_0}M$ and hence $\frak F/\frak D_0$ is its orthonormal frame bundle.
Finally we have the commutative diagram:
$$\CD\frak F/\frak D_0 @>\bar i >>\frak M\times _{\frak D_0}LM^+\\ @V\bar
\pi _{\frak F}VV @V\bar\pi VV \\ \frak M\times _{\frak D_0}M @= \frak M
\times _{\frak D_0}M\endCD\tag2.20$$
where $\bar i$ is the induced embedding.\par
In analogy with Ref. 13 we call $\frak M\times _{\frak D_0}LM^+$
respectively
its reduction $\frak F/\frak D_0$ the universal bundle for gravity.\par
In the remainder of this section we want to define a canonical connection
on the universal bundle. \par
Firstly
there is a natural Riemannian structure on $\frak M$. Each $g\in\frak M$
induces a metric $(,)_g$ on $S^2T^{\ast}M$. For $s_g^1,s_g^2
\in T_g\frak M
\cong\Gamma (S^2T^{\ast}M)$ we can define an inner product on $\frak M$ by
$$<s_g^1,s_g^2>_g:=\int _M(s_g^1,s_g^2)_g\mu _g=\int _M(\hat g^{ij}\hat g
^{kl}(\hat s_g^1)_{ik}(\hat s_g^2)_{jl})\mu _g,\tag2.21$$
where $\mu _g$ is the volume form on $M$ determined by $\hat g$ and the
components of $g$ and $\hat s_g^1$, $\hat s_g^2$ are taken according to
(2.6). Notice, that the integrand on the right hand side is $GL^+(n)$
invariant and thus projects to a well defined function on $M$.\par
It was $\text{shown }^{{}25}$
that $\frak D_0$ acts by isometry, i.e. $<,>$ is $\frak D_0$
invariant. In order to determine the vertical subbundle in $\frak M
\rightarrow\frak M/\frak D_0$, let $R_g^{\frak M}\colon\frak D_0\rightarrow
\frak M$, $R_g^{\frak M}(\phi )=\phi ^{\ast}g$ be the orbit map, then
$$Z_{\zeta}^{\frak M}(g)=T_{id}R_g^{\frak M}(\zeta )=\frac{d}{dt}
\vert _{t=0}\ R_g^{\frak M}
(Fl_t^{\zeta})=\frac{d}{dt}\vert _{t=0}\ (Fl_t^{\zeta})^{\ast}g=L_{\zeta}g
\tag2.22$$
is the fundamental vector field generated by $\zeta=\frac{d}{dt}\vert _{t=0}
Fl_t^{\zeta}\in\frak X(M)$. Here $L_{\zeta}g$ is the Lie derivative of $g$
with respect to $\zeta$. Equivalently, one has
$\widehat{L_{\zeta}g}=L_{Y^{\zeta}}\hat g$
if regarded as element in $C(LM^+,T^{0,2})$. It has been shown in Ref. 25
that the distribution
$$g\mapsto H_g(\frak M)=\lbrace s_g\in\Gamma (S^2T^{\ast}M)\vert \ <s_g,
L_{\zeta}g>_g=0,\qquad\forall\zeta\in\frak X(M)\rbrace\tag2.23$$
defines a connection in $\frak M\rightarrow\frak M/\frak D_0$. Using (2.8)
one finds
$$<s_g,L_{\zeta}g>_g=-\int _M\hat g^{jk}(\hat s_g)_{ij;k}\ \hat\zeta ^i\
\mu _g=<div_gs_g,\zeta >_g,\tag2.24$$
where $\hat g^{jk}(\hat s_g)_{ij;k}$ are the components of the divergence
of $s_g$ with respect to $\Gamma (g)$. Thus the
horizontal bundle consists of all
divergence free, covariant symmetric tensor fields of second degree on $M$.
\par
The corresponding connection one form $\beta\in\Omega ^1(\frak M,\frak X(M))$
is given by
$$\beta _g(s_g)=G_g\circ div_g(s_g),\qquad s_g\in T_g\frak M,\tag2.25$$
where $G_g:=(div_g\circ L_.g)^{-1}$. Here $L_.g$ denotes the mapping
$\frak X(M)\rightarrow\Gamma (S^2T^{\ast}M)$. Together with
the adjoint action of $\frak D$ on $\frak X(M)$
$$(ad(\phi )\zeta )(x):=T_{\phi ^{-1}(x)}\phi\ (\zeta _{\phi ^{-1}(x)}),
\qquad\zeta\in\frak X(M),\ x\in M\tag2.26$$
the transformation property of $\beta$ reads
$$(R_{\phi}^{\frak M\ \ast}\beta )_g(s_g)(x)=\beta _{\phi ^{\ast}g}(\phi ^{
\ast}s_g)(x)=T_{\phi (x)}\phi ^{-1}\ (\beta _g(s_g)_{\phi (x)}),\ \phi\in
\frak D_0,\ x\in M.\tag2.27$$
We now consider the curvature of $\beta$. For vector fields $s^1$, $s^2$ on
$\frak M$ which are horizontal, i.e. $s^1,s^2\in ker(\beta )$, it follows
that
$$\Cal F_g(s_g^1,s_g^2)=(d_{\frak M}\beta +\frac{1}{2}[\beta ,\beta ])_g
(s_g^1,s_g^2)=-\beta _g([s^1,s^2]_g)=-G_g\circ div_g[s^1,s^2]_g.\tag2.28$$
For any $\tau\in\Gamma (S^2T^{\ast}M)$ let $Fl_t^{\tau}(g)=g+t\tau$ denote
the flow on $\frak M$ generated by $\tau$. Then $d_{\Gamma (Fl_t^{\tau}(g))}
\widehat{Fl_t^{\tau}(g)}=0$, $\forall t$, $\text{implies}^{{}23}$
$$d_{\Gamma (g)}\tau =-\Xi (\tau )\cdot\hat g,\tag2.29$$
where the components of $\Xi _g(\tau )\in\Omega _{eq,h}^1(LM^+,T^{1,1})
\cong C(LM^+,T^{1,2})$ are given by the formula
$$\Xi _g(\tau )_{ij}^k:=\frac{d}{dt}\vert _{t=0}(\Gamma (g+t\tau ))_
{ij}^k=
\frac{1}{2}\hat g^{kl}(\hat\tau _{lj;i}+\hat\tau _{il;j}-\hat\tau _{ij;l}).
\tag2.30$$
Notice that an
isomorphism $T^{1,1}\cong gl(n)$ is given by the map $A\rightarrow A^{
\prime}$, where $A^{\prime}(a,v):=a(Av)$ with $A\in gl(n)$, $a\in (\Bbb R^n)
^{\ast}$ and $v\in\Bbb R^n$. Using the following
identity for horizontal vector fields $s^1,s^2$ on $\frak M$
$$\frac{d}{dt}\vert _{t=0}(\hat s_{g+ts^2}^1)_{ij;k}=\Xi _g(s_g^2)_{ki}
^l(\hat s_g^1)_{lj}-\Xi _g(s_g^1)_{kj}^l(\hat s_g^1)_{il},\tag2.31$$
we finally obtain
$$([s^1,s^2]_g)_{i\ ;j}^j=(\hat s_g^1)_{jk;i}(\hat s_g^2)^{jk}-(\hat s_g^2)_
{jk;i}(\hat s_g^1)^{jk}+(\hat s_g^2)_i^k(\hat s_g^1)_{j\ ,k}^j-(\hat s_g^1)_
i^k(\hat s_g^2)_{j\ ,k}^j.\tag2.32$$
If we regard $[s^1,s^2]_{i\ ;j}^j$ as an element of $\Omega _{eq,h}^1(LM^+)
\cong C(LM^+,T^{0,1})$ then the curvature of $\beta$ reads
$$\split\Cal F_g(s_g^1,s_g^2)= & G_g\lbrace <d_{\Gamma (g)}
\hat s_g^2,\hat s_g^1>_g-
<d_{\Gamma (g)}\hat s_g^1,\hat s_g^2>_g+<s_g^1,d_{LM^+}\ tr(\hat s_g^2)>_g
\\ &-<s_g^2,d_{LM^+}\ tr(\hat s_g^1)>_g\rbrace ,\endsplit\tag2.33$$
where $<,>_g$ is the induced inner product in $\Omega (LM^+,T^{p,q})$.\par
Now let $\kappa\colon gl(n)\times gl(n)\rightarrow\Bbb R$, $\kappa (A,B):=
tr(A^tB)$ be an inner product on $gl(n)$, where $A^t$ denotes the
transpose of $A$.
There is a natural Riemann structure on $\frak M\times LM^+$ given
by
$$\mu _{(g,u)}((s_g^1,X_u^1),(s_g^2,X_u^2)):=<s_g^1,s_g^2>_g+(\pi ^{
\ast}g)_u(X_u^1,X_u^2)+\kappa (\Gamma (g)_u(X_u^1),\Gamma (g)_u
(X_u^2)),\tag2.34$$
which induces a Riemannian structure $\mu _0$ on $\frak F$ by
$\mu _0=\bar\chi ^{\ast}\mu$ on $\frak F$. This
inner product is invariant by $SO(n)$ but not under $GL^+(n)$
transformations in general. Moreover, $\mu$ is $\frak D$ invariant because
for $\phi\in\frak D$ and $(s_g^i,X_u^i)\in T_{(g,u)}(\frak M\times LM^+)$,
$i=1,2$, we find
$$\split (w_{\phi}^{\ast}\mu )_{(g,u)} &((s_g^1,X_u^1),(s_g^2,X_u^2))\\ &=
<\phi ^{\ast}s_g^1,\phi ^{\ast}s_g^2>_{\phi ^{\ast}g}+(\pi ^{\ast}\phi
^{\ast}g)_{l(\phi ^{-1})u}(T_ul(\phi ^{-1})X_u^1,T_ul(\phi ^{-1})X_u^2)\\
&\ \ \ +\kappa (\Gamma (\phi ^{\ast}g)_{l(\phi ^{-1})u}T_ul(\phi
^{-1})X_u^1,
\Gamma (\phi ^{\ast}g)_{l(\phi ^{-1})u}T_ul(\phi ^{-1})X_u^2))\\ &=
<s_g^1,
s_g^2>_g+(\pi ^{\ast}g)_u(X_u^1,X_u^2)+\kappa (\Gamma (g)_u(X_u^1),
\Gamma (g)_u(X_u^2)),\endsplit\tag2.35$$
where the invariance of $<,>_g$ has been used. Thus $\mu$ induces a
connection in $\frak M\times LM^+ @>\bar q >>\frak M\times _{\frak D_0}LM^+$
and moreover gives rise to a Riemannian structure $\bar\mu$ on $\frak M
\times _{\frak D_0}LM^+$. Then $\bar\mu _0:=\bar\chi ^{\prime\ \ast}\bar\mu$
is the corresponding metric on $\frak F/\frak D_0$.\par
The vertical subbundle $V^{\bar q}(\frak M\times LM^+)$
of $\frak M\times LM^+ @>\bar q>> \frak M\times _{\frak D_0}
LM^+$ is generated by fundamental vector fields with
respect to the action $w$. In fact, let $\tilde w_{(g,u)}\colon\frak D_0
\rightarrow\frak M\times LM^+$ denote its orbit map, then
$$T_{id}\tilde w_{(g,u)}(\zeta )=\frac{d}{dt}\vert _{t=0}\ w_{Fl_t^{\zeta}}(
g,u)=(Z_{\zeta}^{\frak M}(g),-Y_u^{\zeta}).\tag2.36$$
So we have
$$V^{\bar q}(\frak M\times LM^+)=\bigcup _{(g,u)\in\frak M\times LM^+}
\lbrace (L_{Y^{\zeta}}\hat g,-Y_u^{\zeta})\vert\ \zeta\in\frak X(M)\rbrace
\tag2.37$$
and furthermore $V^{\bar q}(\frak M\times LM^+)\cong V^{\bar p}(\frak F)$,
where the bundle $V^{\bar p}(\frak F)$ is the vertical subbundle of the
principal $\frak D_0$ bundle $\frak F @>\bar p>> \frak F/\frak D_0$. \par
The vertical bundle corresponding to $\frak M\times\widetilde{LM}^+
@>\tilde q>>\frak M\times _{\frak D_0}\widetilde{LM}^+$ is given by
$$V^{\tilde q}(\frak M\times\widetilde{LM}^+)=\bigcup _{(g,[u])\in\frak M
\times\widetilde{LM}^+}\lbrace (L_{Y^{\zeta}}
\hat g,-T_u\tilde\pi\ (Y_u^{\zeta})\vert\ \zeta\in\frak X(M)\rbrace
\tag2.38$$
and finally the vertical bundle of $\frak M\times M@>p>>\frak M
\times _{\frak D_0} M$ reads
$$V^p(\frak M\times M)=\bigcup _{(g,x)\in\frak M\times M}\lbrace
(L_{Y^{\zeta}}\hat g,-\zeta _x)\vert\ \zeta\in\frak X(M)\rbrace .\tag2.39$$
Let us denote by $s_g^h$ the horizontal component of $s_g\in T_g\frak M$
with respect to the connection $\beta$ in (2.25). Then any $(s_g,X_u)\in
T_{(g,u)}(\frak M\times LM^+)$ can be written in the form
$$(s_g,X_u)=(s_g^h,X_u+Y_u^{\beta _g(s_g)})+ (L_{Y^{\beta _g(s_g)}}
\hat g,-Y_u^{\beta _g(s_g)}).\tag2.40$$
\proclaim{Proposition 1} The projector $S\colon T(\frak M\times LM^+)
\rightarrow T(\frak M\times LM^+)$ given by
$$S_{(g,u)}(s_g,X_u):=(s_g^h,X_u+Y_u^{\beta _g(s_g)})\tag2.41$$
defines a connection in $\frak M\times LM^+@>\bar q>>\frak M
\times _{\frak D_0}LM^+$.\endproclaim
\demo{Proof} Let us define $H_{(g,u)}(\frak M\times LM^+):=im(S_{(g,u)})$,
then (2.40) corresponds to the splitting
$$T_{(g,u)}(\frak M\times LM^+)=H_{(g,u)}(\frak M\times LM^+)\oplus V_{(g,u)}
^{\bar q}(\frak M\times LM^+).\tag2.42$$
Since
$$ T_ul(\phi ^{-1})Y_u^{\beta _g(s_g)}=
\frac{d}{dt}\vert _{t=0}\ l(Fl_t^{ad(\phi ^{-1})\beta _g(s_g)})l(\phi ^{-1})
u=Y_{l(\phi ^{-1})u}^{\beta _{\phi ^{\ast}g}(\phi ^{\ast}s_g)}
\tag2.43$$
we finally get
$$\split T_{(g,u)}w_{\phi}\circ S_{(g,u)}(s_g,X_u) &=(\phi ^{\ast}(s_g^h),
T_ul(\phi ^{-1})(X_u+Y_u^{\beta _g(s_g)})\\ &= ((\phi ^{\ast}s_g)^h,T_ul(
\phi ^{-1})X_u+Y_{l(\phi ^{-1})u}^{\beta _{\phi ^{\ast}g}(\phi ^{\ast}s_g)})
\\ &=
S_{w_{\phi}(g,u)}\circ T_{(g,u)}w_{\phi}(s_g,X_u).\endsplit\tag2.44$$
\qed\enddemo
We should remark that $S$ also gives a connection in $\frak F @>>>\frak F/
\frak D_0$ because
$$s_g^h=s_g-L_{Y^{\beta _g(s_g)}}\hat g=-(d_{LM^+}\ \hat g)(X_u+
Y_u^{\beta _g(s_g)})\tag2.45$$
if $(s_g,X_u)\in T_{(g,u)}\frak F$ and hence $S_{(g,u)}(s_g,X_u)\in T_{(g,u)}
\frak F$.\par
The induced metric $\bar\mu$ on $\frak M\times _{\frak D_0}LM^+$ can now be
calculated by the formula
$$\bar\mu _{[g,u]}(T_{(g,u)}\bar q(s_g^1,X_u^1),T_{(g,u)}\bar q(s_g^2,X_u^2))
:=\mu _{(g,u)}(S_{(g,u)}(s_g^1,X_u^1),S_{(g,u)}(s_g^1,X_u^1)).\tag2.46$$
Finally the induced $SO(n)$ action on $T(\frak M\times _{\frak D_0}LM^+)$
is given by
$$r_O^{\prime}(T_{(g,u)}\bar q(s_g,X_u))=T_{(g,u)}\bar q\circ S_{(g,u
\cdot O)}(s_g,T_ur_O(X_u)),\qquad O\in SO(n).\tag2.47$$
\proclaim{Proposition 2} The metric $\bar\mu$ is $SO(n)$ invariant. Thus it
induces a connection in the bundle $\frak M\times _{\frak D_0}LM^+
@>\tilde\pi ^{\prime}>>\frak M\times _{\frak D_0}\widetilde{LM}^+$.
\endproclaim
\demo{Proof} Since $\mu$ is $SO(n)$ invariant we get
$$\split ( &r_O^{\prime\ \ast}\bar\mu )_{[g,u]}(T_{(g,u)}\bar q(s_g^1,X_u^1),
T_{(g,u)}\bar q(s_g^2,X_u^2))\\ &=\bar\mu _{[g,u\cdot O]}(T_{(g,u)}\bar q
\circ S_{(g,u\cdot O)}(s_g^1,T_ur_O(X_u^1)),T_{(g,u)}\bar q
\circ S_{(g,u\cdot O)}(s_g^2,T_ur_O(X_u^2)))\\ &=\mu _{(g,u\cdot O)}(S_
{(g,u\cdot O)}(s_g^1,T_ur_O(X_u^1)),S_{(g,u\cdot O)}(s_g^2,T_ur_O(X_u^2)))\\
&=(r_O^{\ast}\mu )_{(g,u)}(S_{(g,u)}(s_g^1,X_u^1),
S_{(g,u)}(s_g^2,X_u^2))\\
&=\bar\mu _{[g,u]}(T_{(g,u)}\bar q(s_g^1,X_u^1),
T_{(g,u)}\bar q(s_g^2,X_u^2)).\endsplit\tag2.48$$
\qed\enddemo
As a consequence, $\bar\mu _0$ induces a connection in the universal bundle.
It is our aim to determine this connection explicitly.\par
Since $\bar q^{\ast}T(\frak M\times _{\frak D_0}LM^+)\cong \bigcup _{(g,u)
\in\frak M\times LM^+}H_{(g,u)}(\frak M\times LM^+)$ we can identify any
tangent vector $\tau _{[g,u]}$ of $\frak M\times _{\frak D_0}LM^+$ with a
vector $(s_g,X_u)$ which is horizontal with respect to $S$. \par
Let $\tau _{[g,u]}=T_{(g,u)}\bar q(s_g,X_u)$ where $(s_g,X_u)\in H_{(g,u)}(
\frak M\times LM^+)$. For $\tau$ being vertical in $\frak M\times _{
\frak D_0}LM^+\rightarrow\frak M\times _{\frak D_0}\widetilde{LM}^+$ it
follows that
$$T_{(g,\tilde\pi (u))}q\ (s_g,T_u\tilde\pi (X_u))=0,\tag2.49$$
where $q\colon\frak M\times\widetilde{LM}^+\rightarrow\frak M
\times _{\frak D_0}\widetilde{LM}^+$ is the corresponding projection.\par
This implies that $(s_g,T_u\tilde\pi (X_u))\in V_{(g,[u])}^q(\frak M\times
\widetilde{LM}^+)$. Therefore $s_g=T_u\tilde\pi (X_u))=0$. So there exists
$\xi
\in so(n)$ such that $X_u=Z_{\xi}(u)$, where $Z_{\xi}$ is the fundamental
vector field on $LM^+$ generated by $\xi$. As a result, the vertical bundle
is
$$V^{\tilde\pi ^{\prime}}(\frak M\times _{\frak D_0}LM^+)=\bigcup _{[g,u]\in
\frak M\times _{\frak D_0}LM^+}\lbrace T_{(g,u)}\bar q(0_g,Z_{\xi}(u))\vert
\ \xi\in so(n)\rbrace .\tag2.50$$
Now we are able to determine the horizontal bundle $H^{\bar\mu}(\frak M
\times _{\frak D_0}LM^+)$ of the connection which is induced by the metric
$\bar\mu$. If $\tau _{[g,u]}\in H_{[g,u]}^{\bar\mu}(\frak M\times _{
\frak D_0}LM^+)$, then it must fulfill the following equation
$$\bar\mu _{[g,u]}(T_{(g,u)}\bar q(0_g,Z_{\xi}(u)),T_{(g,u)}\bar q(s_g,X_u))=
\kappa (\xi,\Gamma (g)_u(X_u))=0\tag2.51$$
for all $\xi\in so(n)$. Here we have represented $\tau _{[g,u]}=T_{(g,u)}
\bar q(s_g,X_u)$, with $(s_g,X_u)\in H_{(g,u)}(\frak M\times LM^+)$.
Since $\Gamma (g)$ admits a natural decomposition
$$\Gamma (g)=\Gamma ^a(g)+\Gamma ^{s}(g):=\frac{1}{2}(\Gamma (g)-
\Gamma (g)^t)+\frac{1}{2}(\Gamma (g)+\Gamma (g)^t)\tag2.52$$
into an antisymmetric and a symmetric part with respect to the standard
scalar product in $\Bbb R^n$, (2.51) implies that
$$\kappa (\xi,\Gamma ^a(g)_u(X_u))=0,\qquad\forall\xi\in so(n).\tag2.53$$
But $\Gamma ^a(g)$ gives a connection in $LM^+\rightarrow\widetilde{LM}^+$
whose
horizontal bundle will be denoted by $H^{\Gamma ^a}(LM^+)$. So the horizontal
bundle we are looking for is given by
$$H^{\bar\mu}(\frak M\times _{\frak D_0}LM^+)=\bigcup _{[g,u]\in
\frak M\times _{\frak D_0}LM^+}\lbrace T_{(g,u)}\bar q(s_g,X_u)\vert\ s_g\in
H_g(\frak M), X_u\in H_u^{\Gamma ^a}(LM^+)\rbrace .\tag2.54$$
Let us now define the following connection one form on
$\frak M\times LM^+\rightarrow\frak M\times\widetilde{LM}^+$
$$\epsilon _{(g,u)}(s_g,X_u)=\epsilon _{(g,u)}^{(1,0)}(s_g)+\epsilon _{(g,u)}
^{(0,1)}(X_u)=(i_{Y^{\beta _g(s_g)}}\Gamma ^a(g))(u)+\Gamma ^a(g)_u(X_u).
\tag2.55$$
\proclaim{Proposition 3} The connection $\epsilon$ induces a well defined
connection $\bar
\epsilon$ in $\frak M\times _{\frak D_0}LM^+\rightarrow\frak M\times _
{\frak D_0}\widetilde{LM}^+$. In fact, $\bar\epsilon$ coincides
with the connection induced by $\bar\mu$.\endproclaim
\demo{Proof} $\epsilon$ is $\frak D_0$ invariant since
$$\split &(w_{\phi}^{\ast}\epsilon )_{(g,u)}(s_g,X_u)=
\epsilon _{(\phi ^{\ast}g,l(\phi ^{-1})u)}(T_gR_{\phi}^{\frak M}s_g,T_ul(
\phi ^{-1})X_u)\\ &=\Gamma ^a(\phi^{\ast}g)_{l(\phi ^{-1})u}(T_ul(\phi ^{-1})
X_u)+\Gamma ^a(\phi^{\ast}g)_{l(\phi ^{-1})u}(Y_{l(\phi ^{-1})u}^{\beta _{
\phi ^{\ast}g}(\phi ^{\ast}s_g)})\\ &=\epsilon _{(g,u)}(s_g,X_u).\endsplit
\tag2.56$$
Furthermore we have $\epsilon _{(g,u)}(Z_{\zeta}^{\frak M}(g),-Y_u^{
\zeta})=0$ implying that $\epsilon$ is basic. Thus there exists a one form
$\bar\epsilon\in\Omega ^1(\frak M
\times _{\frak D_0}LM^+,so(n))$ such that $\bar q^{\ast}\bar\epsilon =
\epsilon$.
Let $\tau _{[g,u]}=T_{(g,u)}\bar q(s_g,X_u)$ be horizontal with respect to
$\bar\epsilon$. Because $(s_g,X_u)\in H_{(g,u)}(\frak M\times LM^+)$ we
conclude from
$$\bar\epsilon _{[g,u]}(\tau _{[g,u]})=\epsilon _{(g,u)}(s_g,X_u)=\Gamma ^a
(g)_u(X_u)=0\tag2.57$$
that $X_u$ must be horizontal with respect to $\Gamma ^a(g)$.\qed\enddemo
In fact, $\bar\chi ^{\prime\ \ast}\bar\epsilon$ agrees with the connection
induced by the metric $\bar\chi ^{\prime\ \ast}\bar\mu$ on $\frak F/
\frak D_0$. In analogy with ref. 13
this connection will be called the universal connection.\par
For our purpose it is, however, necessary to have a connection on $\frak M
\times _{\frak D_0}LM^+$ which is reducible to a connection on $\frak F/
\frak D_0$, i.e.
a connection which is compatible with the vertical metric $\bar\nu$.\par
Therefore let us define the following connection
on $\frak M\times LM^+\rightarrow\frak M\times M$ by
$$\omega _{(g,u)}(s_g,X_u)=\omega _{(g,u)}^{(1,0)}(s_g)+
\omega _{(g,u)}^{(0,1)}(X_u)=\Gamma (g)_u(Y_u^{
\beta _g(s_g)})+\rho _g(s_g^h)(u)+\Gamma (g)_u(X_u).\tag2.58$$
Here $\rho$ is a $C(LM^+,T^{1,1})$ valued one form on $\frak M$ defined
by the components according to (2.6)
$$\rho _g(s_g)_{\ k}^i(u)=\frac{1}{2}\ \hat g^{ij}(u)(\hat s_g)_{jk}(u),
\tag2.59$$
where it is summed over the index j. Notice, that $\hat g^{ij}$ are the
components of $\hat g^{-1}$.
\proclaim{Proposition 4} The connection $\omega$ induces a connection
$\bar\omega$ on $\frak M\times _{\frak D_0}LM^+\rightarrow \frak M\times _{
\frak D_0}M$. Furthermore $\omega$ and $\bar\omega$
satisfy $i^{\ast}\omega=\bar\chi ^{\ast}
\epsilon$ and $\bar i^{\ast}\bar\omega =\bar
\chi ^{\prime\ \ast}\bar\epsilon$.\endproclaim
\demo{Proof} It is easy to verify that $\omega$ is basic, thus inducing a
connection $\bar\omega$. For $(s_g,X_u)\in T_{(g,u)}\frak F$ one finds
from (2.7)
$$(\hat s_g)_{ij}(u)=-d_{LM^+}\hat g_{ij}(X_u)=-\Gamma (g)_u(X_u)_{\ j}^k
\hat g_{ki}(u)-\Gamma (g)_u(X_u)_{\ i}^k\hat g_{kj}(u).\tag2.60$$
Together with $s_g=s_g^h+L_{Y^{\beta _g(s_g)}}\hat g$ and $\hat g^{ij}(u)=
\delta ^{ij}$ one obtains
$$(i^{\ast}\omega )_{(g,u)}(s_g,X_u)_{\ k}^i=\Gamma ^a(g)_u(X_u)+\Gamma
^a(g)_u(Y_u^{\beta _g(s_g)})=(\bar\chi ^{\ast}\epsilon )_{(g,u)}(s_g,X_u),
\tag2.61$$
and hence $\bar i^{\ast}\bar\omega =\bar\chi ^{\prime\ \ast}\bar\epsilon$.
\qed\enddemo
Let $(f^{t_g})_{\ k}^i:=\hat g^{ij}f_{\ j}^l\hat g_{lk}$ denote the
transpose of $f\in C(LM^+,T^{1,1})$ with respect to the metric $g$ and
let $f^{a_g}:=1/2(f-f^{t_g})$ be its antisymmetric part. Furthermore let
$R(g)$ be the curvature of the Levi Civita connection. According to the
product structure of $\frak M\times LM^+$ there is a bigrading of the space
of differential forms $\Omega (\frak M\times LM^+)=\oplus _{i,j\geq 0}
\Omega ^{(i,j)}$ on it. The exterior derivative splits into $d=d_{\frak M}+
\hat d_{LM^+}$ where
$$\split & d_{\frak M}\colon\Omega ^{(i,j)}\rightarrow\Omega ^{(i+1,j)}\\
& \hat d_{LM^+}\colon\Omega ^{(i,j)}\rightarrow\Omega ^{(i,j+1)},\qquad
\hat d_{LM^+}:=(-1)^id_{LM^+}.\endsplit\tag2.62$$
\proclaim{Proposition 5} The components of the curvature $\Omega _{\omega}$
of $\omega$ are given by
$$\split\Omega _{\omega\
(g,u)}^{\ \ (2,0)}(s_g^1,s_g^2)= & (i_{Y^{\Cal F_g(s_g^1,s_g^2)}}
\Gamma ^{a_g}(g)
+i_{Y^{\beta _g(s_g^2)}}\Xi _g^{a_g}(s_g^1)-i_{Y^{\beta _g(s_g^1)}}\Xi _g
^{a_g}(s_g^2)
\\ &+[i_{Y^{\beta _g(s_g^1)}}\Gamma ^{a_g}(g),i_{Y^{\beta _g(s_g^2)}}
\Gamma ^{a_g}(g)]-
i_{[Y^{\beta _g(s_g^1)},Y^{\beta _g(s_g^2)}]}\Gamma ^{a_g}(g)\\
& -[\rho _g(s_g^1),\rho _g(s_g^2)]+[\rho _g(s_g^1),i_{Y^{\beta _g(s_g^2)}}
\Gamma ^{t_g}(g)]\\ &-[\rho _g(s_g^2),i_{Y^{\beta _g(s_g^1)}}
\Gamma ^{t_g}(g)])(u)\\ \Omega _{\omega\ (g,u)}^{\ \ (1,1)}(s_g,X_u)= &
(\Xi _g^{a_g}(s_g)-d_{\Gamma (g)}(i_{Y^{\beta _g(s_g)}}\Gamma ^{a_g}(g))_u(
X_u)\\
\Omega _{\omega\ (g,u)}^{\ \ (0,2)}(X_u^1,X_u^2) =& R(g)_u(X_u^1,X_u^2)
\endsplit\tag2.63$$\endproclaim
\demo{Proof} Since $s_g^h=s_g-L_{Y^{\beta _g(s_g)}}\hat g$, one can write
(2.58) in the form
$$\omega _{(g,u)}(s_g,X_u)=\Gamma ^{a_g}(g)_u(Y_u^{\beta _g(s_g)})+\rho _g(
s_g)(u)+\Gamma (g)_u(X_u).\tag2.64$$
The result for the (0,2) component is obvious. The (1,1) component can be
calculated from the formula
$$\Omega _{\omega}^{(1,1)}=d_{\frak M}\ \omega ^{(0,1)}-d_{LM^+}\
\omega ^{(1,0)}-[\omega ^{(0,1)},\omega ^{(1,0)}].\tag2.65$$
Using (2.29) we get
$$d_{\Gamma (g)}\rho _g(s_g)_{\ k}^i=\frac{1}{2}\ (\Xi _g(s_g)_{\ k}^i+
\Xi _g^{t_g}(s_g)_{\ k}^i)\tag2.66$$
and hence inserting into (2.65) gives the final result. In order to find
the (2,0) component we observe that the following identity holds for
any vector fields $s^1,s^2$ on $\frak M$
$$d_{\frak M}(i_{Y^{\beta (s^2)}}\Gamma ^{a_g}(g))(s_g^1)
=\Xi _g^{a_g}(s_g^1)_u(Y_u^{\beta _g(s_g^2)})+\Gamma ^{a_g}(g)_u(Y_u^{
s_g^1(\beta (s^2))})+[\rho _g(s_g^1),i_{Y^{\beta _g(s_g^2)}}\Gamma ^{t_g}
(g)].\tag2.67$$
Using (2.28) the result follows.\qed\enddemo
Now the curvature $\Omega _{\bar\omega}$ of $\bar\omega$ can be easily
calculated by noticing that
$$\Omega _{\bar\omega\ [g,u]}(\tau _{[g,u]}^1,\tau _{[g,u]}^2)=\Omega _{
\omega\ (g,u)}((s_g^1,X_u^1),(s_g^2,X_u^2)),\tag2.68$$
where $\tau _{[g,u]}^i=T_{(g,u)}\bar q(s_g^i,X_u^i)$ for $i=1,2$ and
$(s_g^i,X_u^i)\in H_{(g,u)}(\frak M\times LM^+)$. Hence the relevant
components are given by
$$\split &\Omega _{\omega\ (g,u)}^{\ \ (2,0)}(s_g^1,s_g^2)=(i_{Y^{\Cal
F_g(s_g^1,s_g^2)}}\Gamma ^{a_g}(g)-[\rho _g(s_g^1),\rho _g(s_g^2)])(u)\\
&\Omega _{\omega\ (g,u)}^{\ \ (1,1)}(s_g,X_u)=\Xi _g^{a_g}(s_g)_u(X_u)\\
&\Omega _{\omega\ (g,u)}^{\ \ (0,2)}(X_u^1,X_u^2)=R(g)_u(X_u^1,X_u^2).
\endsplit\tag2.69$$
\bigskip\bigskip
{\bf III. Consistent gravitational anomalies}
\bigskip
A. The gravitational BRS multiplet
\bigskip
In this section we shall construct an appropriate geometrical framework to
dicuss both the consistent Lorentz and diffeomorphism anomalies. It will be
explicitly proved that the consistent gravitational anomalies can be derived
using the local index formula of Bismut and $\text{Freed.}^{{}11}$ \par
Let $\Cal G$ denote the group defined by
$$\Cal G=\lbrace F\in Aut_0(LM^+)\vert F(u)=u\cdot\hat F(u),\ \hat F(u)\in
SO(n),
\ \forall u\in LM^+\rbrace .\tag3.1$$
For a fixed metric $g$, the group $\Cal G$ can be identified with the gauge
group of $SO_g(M)$.\par
The mappings $\bar{ev}(F,g,u)=(g,F(u))$ and $\varrho (F,g,x)=
(F,ev(g,x))$ make the following diagram commute
$$\CD\Cal G\times\frak F @>\bar\varrho >> \Cal G\times\frak M\times
LM^+ @>\bar{ev}>> \frak M\times LM^+\\ @V id_{\Cal G}\times\pi _{\frak F}VV
@V id_{\Cal G}\times\pi VV @V \pi VV \\ \Cal G\times
\frak M\times M @>\varrho >> \Cal G\times\frak M\times\widetilde{LM}^+
@>pr >>\frak M\times\widetilde{LM}^+\endCD\tag3.2$$
where $\bar\varrho$ is the relevant bundle map and $pr$ is the canonical
projection. Thus we find the bundle isomorphism
$$\Cal G\times\frak F\cong \varrho ^{\ast}(\Cal G\times\frak M\times LM^+).
\tag3.3$$
Let $\Theta\in\Omega ^1(\Cal G,Lie\Cal G)$ denote the Maurer Cartan form on
$\Cal G$.
Any $\Cal Y_F\in T_F\Cal G$ can be written in the form $\Cal Y_F=\Cal Y_{
\Theta _F(\Cal Y_F)}^{\text{left}}(F)$, where $\Cal Y_{\Theta _F(\Cal Y_F)}^
{\text{left}}$ denotes the left invariant vector field on $\Cal G$ induced
by $\Theta _F(\Cal Y_F)\in Lie\Cal G$. Hence $\Cal Y$ generates the flow
$Fl_t^{\Cal Y}(F)(u)=F(u)\cdot exp\ t\Theta _F(\Cal Y_F)(u)$. \par
Let us
consider the connection $\tilde\omega :=\bar{ev}^{\ast}\omega$ on
$\Cal G\times\frak M\times LM^+\rightarrow\Cal G\times\frak M\times M$.
Since
$$T_{(F,g,u)}\bar{ev}(\Cal Y_F,s_g,X_u))=(s_g,T_uF\ (X_u+
Z_{\Theta _F(\Cal Y_F)}(u))),\tag3.4$$
where $Z_{\Theta _F(\Cal Y_F)}$ is the fundamental vector field on
$LM^+$ the components of $\tilde\omega $ are given by
$$\split\tilde\omega _{(F,g,u)}(\Cal Y_F,s_g,X_u) &=\tilde\omega _{(F,g,u)}
^{(1,0,0)}(\Cal Y_F)+\tilde\omega _{(F,g,u)}^{(0,1,0)}(s_g)+\tilde
\omega _{(F,g,u)}^{(0,0,1)}(X_u)\\ &= \Theta _F(\Cal Y_F)(u)+F^{\ast}
(i_{Y^{\beta _g(s_g)}}\Gamma (g)+\rho (s_g^h))(u)+(F^{\ast}\Gamma (g))_
u(X_u).\endsplit\tag3.5$$
It is evident that $\tilde\omega $ restricted to $\Cal G\times\frak F$
coincides with $(\bar{ev}\circ\bar\varrho)^{\ast}\epsilon$.\par
We define the semidirect product $\frak G:=\Cal G\rtimes\frak D_0$
with group structure
$$(F_1,\phi _1)\cdot (F_2,\phi _2):=(l(\phi _2^{-1})
\circ F_1\circ l(\phi _2)\circ F_2,\phi _1\circ\phi _2).\tag3.6$$
Its Lie algebra $Lie\frak G=Lie\Cal G\rtimes\frak X(M)$ is the semidirect
product of $\frak X(M)$ and $Lie\Cal G$, where we have the identification
$$Lie\Cal G=\lbrace\xi\colon LM^+\rightarrow so(n)\vert \ \xi (u\cdot O)=
ad(O^{-1})\ \xi (u),\ O\in SO(n)\rbrace .\tag3.7$$
We define a right action of $\frak G$ on $\Cal G\times\frak M\times LM^+$ by
$$\alpha _{(\phi ,F^{\prime})}(F,g,u):=(l(\phi ^{-1})\circ F\circ l(\phi )
\circ
F^{\prime},\phi ^{\ast}g,(l(\phi )\circ F^{\prime})^{-1}(u)).\tag3.8$$
This action is free and naturally extends to a free action on $\Cal G\times
\frak F$. Furthermore, $\alpha$ commutes with the principal $GL^+(n)$ action
and thus induces a
free action on $\Cal G\times\frak M\times M$ by
$$\bar\alpha _{(\phi ,F^{\prime})}(F,g,x):=(l(\phi ^{-1})\circ F\circ l(
\phi )\circ F^{\prime},\phi ^{\ast}g,\phi ^{-1}(x)).\tag3.9$$
Now we can identify
$(\Cal G\times \frak M\times LM^+)/\frak G\cong\frak M\times _{
\frak D_0}LM^+$ by the map
$$h\colon (\Cal G\times \frak M\times LM^+)/\frak G\rightarrow
\frak M\times _{\frak D_0}LM^+,\ [F,g,u]_{\frak G}\mapsto [g,F(u)]_{
\frak D_0},\tag3.10$$
and furthermore we have $(\Cal G\times\frak M\times M)/\frak G
\cong\frak M\times _{\frak D_0}M$ provided by
$$h^{\prime}\colon (\Cal G\times \frak M\times M)/\frak G\rightarrow
\frak M\times _{\frak D_0}M,\ [F,g,x]_{\frak G}\mapsto [g,x]_{
\frak D_0},\tag3.11$$
where the brackets are the equivalence classes with respect to the actions
of $\frak G$ and $\frak D_0$ respectively. We can summarize our constructions
in the following commutative diagram:
$$\CD \Cal G\times \frak F @>\bar f^{\prime}>> \frak F/\frak D_0 @>\bar i >>
\frak M\times _{\frak D_0}LM^+ @<\bar f<<\Cal G\times\frak M\times LM^+\\
@V id_{\Cal G}\times\pi _{\frak F}VV @V\bar\pi _{\frak F}VV @V\bar\pi VV
@V id_{\Cal G}\times\pi VV\\
\Cal G\times\frak M\times M @>f>>\frak M\times _{\frak D_0}M
@= \frak M\times _{\frak D_0}M @<f<< \Cal G\times\frak M\times M\\
@VVV @VVV @VVV @VVV \\ \Cal G\times\frak M @>\pi _{\Cal G\times\frak M}>>
\frak M/\frak D_0 @= \frak M/\frak D_0 @<\pi _{\Cal G\times\frak M}<< \Cal G
\times\frak M\endCD\tag3.12$$
where $f(F,g,x)=[g,x]_{\frak D_0}$, $\bar f(F,g,u)=[g,F(u)]_{\frak D_0}$
and $\bar f^{\prime}=\bar f\circ i$. Obviously, one has the bundle
isomorphisms
$$f^{\ast}(\frak F/\frak D_0)\cong\Cal G\times\frak F,\qquad
f^{\ast}(\frak M\times _{\frak D_0}LM^+)\cong\Cal G\times\frak M\times
LM^+.\tag3.13$$
\proclaim{Proposition 6} The connection $\tilde\omega$ satisfies $\bar f^{
\ast}\bar\omega =\tilde\omega$ and
descends to a well
defined connection $\tilde\omega ^{\prime}$ on
$(\Cal G\times\frak M\times LM^+)/\frak G\rightarrow (\Cal G\times\frak M
\times M)/\frak G$.\endproclaim
\demo{Proof} Since $\bar f=\bar q\circ\bar{ev}$ the relation between $\bar
\omega$ and $\tilde\omega$ is evident.
Because $\bar{ev}\circ\alpha _{(\phi ,F^{\prime})}=w_{\phi}
\circ\bar{ev}$ we find that $\alpha _{(\phi ,F^{
\prime})}^{\ast}\tilde\omega =\tilde\omega$, $\forall (F^{\prime},\phi )\in
\frak G$.
Let $\bar\alpha _{(F,g,u)}(F^{\prime},\phi ):=\alpha _{(\phi ,F^{\prime})}
(F,g,u)$ be the orbit map corresponding to $\alpha$. Then the fundamental
vector field generated by $(\xi,\zeta )\in Lie\frak G$ reads
$$\bar\zeta _{(F,g,u)}(\xi ,\zeta ):=\frac{d}{dt}\vert _{t=0}\ \bar\alpha _{(
F,g,u)}(exp(t\xi ),Fl_t^{\zeta}).\tag3.14$$
Hence
$\tilde\omega _{(F,g,u)}(\bar\zeta _{(F,g,u)}(\xi ,\zeta ))=
\omega _{(g,F(u))}(Z_{\zeta}^{\frak M}(g),-Y_{F(u)}^{\zeta})=0$,
proving that $\tilde\omega$ is basic. \qed\enddemo
According to the triple grading of
$$\Omega (\Cal G\times\frak M\times LM^+)=\oplus _{i,j,k}\ \Omega ^{(i,j,k)}
(\Cal G\times\frak M\times LM^+)\tag3.15$$
the exterior
derivative can be written in the form $d=d_{\Cal G}+\hat d_{\frak M}+\hat d_
{LM^+}$, where
$$\alignat2 & d_{\Cal G}\colon\Omega ^{(i,j,k)}\rightarrow\Omega ^{(i+1,j,k)}
&& \\
& \hat d_{\frak M}\colon\Omega ^{(i,j,k)}\rightarrow\Omega ^{(i,j+1,k)},
&& \qquad\hat d_{\frak M}:=(-1)^id_{\frak M}\\
& \hat d_{LM^+}\colon\Omega ^{(i,j,k)}\rightarrow\Omega ^{(i,j,k+1)}, &&
\qquad\hat d_{LM^+}:=(-1)^{i+j}d_{LM^+}.\tag3.16\endalignat$$
Using (3.4) we can calculate the curvature $\Omega _{\tilde\omega}$ of
$\tilde\omega$
$$\split &\Omega _{\tilde\omega (F,g,u)}((\Cal Y_F^1,s_g^1,X_u^1),(
\Cal Y_F^2,
s_g^2,X_u^2))=(\bar{ev}^{\ast}\Omega _{\omega})_{(F,g,u)}((\Cal Y_F^1,s_g^1,
X_u^1),
(\Cal Y_F^2,s_g^2,X_u^2))\\ &=\Omega _{\omega\ (g,F(u))}((s_g^1,T_uF(X_u^1),
(s_g^2,T_uF(X_u^2)),\endsplit\tag3.17$$
since $T_uF(Z_{\xi}(u))$ is vertical $\forall\xi\in Lie\Cal G$. Together
with (2.63) one can prove
\proclaim{Proposition 7} The components of $\Omega _{\tilde\omega}$ are
given by
$$\split &\Omega _{\tilde\omega}^{\ (2,0,0)} =\Omega _{\tilde\omega}
^{\ (1,1,0)}=\Omega _{\tilde\omega}^{\ (1,0,1)}=0\\
&\tilde\Omega _{\tilde\omega\ (F,g,u)}^{\ \ \ (0,2,0)}(s_g^1,s_g^2)=
\Omega _{\omega\
(g,F(u))}^{\ \ \ (2,0)}(s_g^1,s_g^2)\\ &\Omega _{\tilde\omega\ (F,g,u)}^{\ \
\ (0,0,2)}(X_u^1,X_u^2)=(F^{\ast}R(g))_u(X_u^1,X_u^2)\\
&\Omega _{\tilde\omega\ (F,g,u)}^{\ \ \ (0,1,1)}(s_g,X_u)=
F^{\ast}(\Xi ^{a_g}(s_g)-d_{\Gamma (g)}(i_{Y^{\beta _g(s_g)}}\Gamma ^{a_g}
(g))_u(X_u).\endsplit\tag3.18$$
\endproclaim
Now we shall display the gravitational BRS multiplet for combined
Lorentz transformations and diffeomorphisms. For fixed $g\in\frak M$ let us
define the map
$$\gather i_g\colon\Cal G\times\frak D_0\times LM^+\hookrightarrow
\Cal G\times\frak M\times LM^+\\ i_g(F,\phi ,u):=(F,\phi ^{\ast}g,u).
\tag3.19\endgather$$
{}From (2.30) one obtains
$$\Xi _g(Z_{\zeta}^{\frak M}(g))=\frac{d}{dt}\vert _{t=0}\Gamma (Fl_t^{
\zeta\ \ast}g)=L_{Y^{\zeta}}\Gamma (g).\tag3.20$$
Using the formula
$$L_X\Gamma (g)=i_XR(g)+d_{\Gamma (g)}(i_X\Gamma (g))\tag3.21$$
which holds for any $X\in\frak X(LM^+)$ the following result can be
established.
\proclaim{Corollary 8} The components of $i_g^{\ast}\Omega _{\tilde\omega}$
are given by
$$\split &(i_g^{\ast}\Omega _{\tilde\omega})^{(2,0,0)}=(i_g^{\ast}
\Omega _{\tilde\omega})^{(1,1,0)}=(i_g^{\ast}\Omega _{\tilde\omega})
^{(1,0,1)}=0\\
& (i_g^{\ast}\Omega _{\tilde\omega})_{(F,\phi ,u)}^{(0,2,0)}(\zeta _{\phi}^1,
\zeta _{\phi}^2)=R(g)_u(Y_u^{c_{\phi}(\zeta _{\phi}^1)},Y_u^{c_{\phi}(
\zeta _{\phi}^2)})\\ & (i_g^{\ast}\Omega _{\tilde\omega})_{(F,\phi ,u)}
^{(0,0,2)}
(X_u^1,X_u^2)=((l(\phi )\circ F)^{\ast}R(g))_u(X_u^1,X_u^2)\\
& (i_g^{\ast}\Omega _{\tilde\omega})_{(F,\phi ,u)}^{(0,1,1)}(\zeta _{\phi},
X_u)=(i_{Y^{c_{\phi}(\zeta _{\phi})}}R(g))_u(X_u),\endsplit\tag3.22$$
where $c\in\Omega ^1(\frak D_0,\frak X(M))$ is the Maurer Cartan form on
$\frak D_0$ and $\zeta _{\phi}^i\in T_{\phi}\frak D_0$ with $i=1,2$.
\endproclaim
In order to derive the BRS structure in our formalism we shall use the
following abbreviations:
$$\alignat2 & (i_g^{\ast}\tilde\omega )^{(1,0,0)} =\Theta &&\qquad
(i_g^{\ast}\Omega _{\tilde\omega})^{(0,1,1)} =i_{c}R\\
& (i_g^{\ast}\tilde\omega )^{(0,0,1)} =\Gamma &&\qquad
(i_g^{\ast}\Omega _{\tilde\omega})^{(0,2,0)} =i_{c}i_{c}R\\
&(i_g^{\ast}\tilde\omega )^{(0,1,0)} =i_{c}\Gamma  &&\qquad
\delta =d_{\Cal G}+\hat d_{\frak D_0}.\tag3.23\endalignat$$
Using (3.22) we finally obtain the system of equations
$$\split &\delta \Gamma =d_{\Gamma}(\Theta +i_{c}\Gamma )+i_{c}R\\
&\delta (\Theta +i_{c}\Gamma )= i_{c}i_{c}R-\frac{1}{2}[\Theta +
i_{c}\Gamma ,\Theta +i_{c}\Gamma ]\\ &\delta c=-\frac{1}{2}[c,c],
\endsplit\tag3.24$$
where the last equation in (3.24) is the Maurer Cartan equation
in $\frak D_0$. In Ref. 26 these equations have been derived in a
different way.
Thus we have explicitly verified that the bundle $\frak G
\times LM^+\rightarrow\frak G\times M$ provides
an appropriate geometrical framework to formulate the gravitational BRS
relations.\par
\bigskip
B. Local index formula and consistent descent equations
\bigskip
The quantum version of a classical action which describes the kinematics of
fields in an external bosonic background is usually represented by a
determinant of a certain family of Dirac operators. Geometrically, the
determinant can be viewed as a section of the determinant line bundle
associated with the given operator family. In the gravitational case the
configuration space and the corresponding Dirac family can be constructed as
follows:\par
The spin structure on $M$
determines a lift of $\Cal G\times\frak F$ to a principal $Spin(n)$ bundle
$\Cal G\times\tilde\frak F$ over $\Cal G\times\frak M\times M$, where $\tilde
\frak F$ is the double cover of $\frak F$. For fixed metric $g$, $\tilde
\frak F$ restricts to the bundle of spin frames $Spin_g(M)$ on $M$. Let
$\Delta ^{\pm}$ denote the half-spin representations. For fixed $(F,g)\in
\Cal G\times\frak M$, the associated
bundles $\Lambda ^{\pm}=(\Cal G\times\tilde\frak F)\times _{Spin(n)}\Delta ^
{\pm}$ restrict to the bundles of positive and negative spinors over $M$
respectively. Consider the metric compatible connection $F^{\ast}\Gamma (g)$
on $LM^+$ where $F\in\Cal G$. For fixed $F$ and metric $g$, it reduces to a
$SO(n)$ connection on $SO_g(M)$ and can be uniquely lifted to a
connection on $Spin_g(M)$. The chiral Dirac operators can now be
constructed in the usual way
$$D_{(F,g)}\colon\Gamma (\Lambda _{\lbrace (F,g)\rbrace
\times M}^{\pm})\rightarrow\Gamma (\Lambda _{\lbrace (F,g)\rbrace\times M}
^{\mp}).\tag3.25$$
They fit together to form a family
of elliptic operators parametrized by $\Cal G\times\frak M$.
If we consider only
those diffeomorphisms from $\frak D_0$ which preserves the spin
$\text{structure}^{{}27}$
then there exists a lifting to an action of $\frak G$ on $\Cal G\times
\tilde\frak F$. Since the operators (3.25) transform under $\frak G$
transformations according to
$$(l(\phi )\circ F^{\prime})^{\ast}\circ D_{(F,g)}\circ (l(\phi )\circ F^{
\prime})^{-1\ \ast}=D_{(l(\phi ^{-1})\circ F\circ l(\phi )\circ F^{\prime},
\phi ^{\ast}g)},\tag3.26$$
$\lbrace D_{(F,g)}\rbrace$ is an equivariant
family. The action of $\frak G$ is free and therefore one can
equivalently work on the quotient family of operators parametrized by
$(\Cal G\times\frak M)/\frak G\cong\frak M/
\frak D_0$. \par
In the following we shall specify the geometrical data according to Ref. 10
in
order to calculate the curvature of the determinant line bundle associated
with (3.25).\par
Let us define a connection in the trivial fibre bundle $\frak M\times M
\rightarrow\frak M$ by
$$P\colon T(\frak M\times M)\rightarrow T(\frak M\times M),\qquad P_{(g,x)}(
s_g,\zeta _x):=(0_g,\zeta _x+\beta _g(s_g)_x).\tag3.27$$
The metric $<,>$ lifted to the horizontal subspaces (determined by P)
together with $\nu$ combine to form a metric $\nu ^{\prime}$ on $\frak M
\times M$. Let $\nabla ^{\nu ^{\prime}}$ denote the corresponding Levi-Civita
connection, then $\nabla ^{\text{ver}}:=P\circ\nabla ^{\nu ^{\prime}}$ is a
connection on $\frak M\times TM$. Explicitly, $\nabla ^{\text{ver}}$ is
determined by
$$\split 2\nu (\nabla _U^{\text{ver}}V,W) & =\nu (P[U,V],W)-\nu (P(U),
[V,W])+\nu (P[W,
U],V)\\ &+U(\nu (V,W))+V(\nu (P(U),W))-W(\nu (P(U),V)),\endsplit\tag3.28$$
where $U\in\frak X(\frak M\times M)$ and $V,W\in\Gamma (\frak M\times TM)$.
With respect to the identification $\frak M\times TM\cong(\frak M\times LM^+)
\times _{GL^+(n)}\Bbb R^n$ every section $V\in\Gamma (\frak M\times TM)$
can be written in the following form
$$V(g,x)=(0_g,\bar V_{(g,x)})=(0_g,[u,\hat V(g,u)]).\tag3.29$$
\proclaim{Proposition 9} The connection $P$ projects to a well defined
connection $\bar P$ on $\frak M\times _{\frak D_0}M\rightarrow\frak M/
\frak D_0$.\endproclaim
\demo{Proof} Since $T(\frak M\times _{\frak D_0}M)\cong T\frak M
\times _{T\frak D_0}TM$, $T_{(g,x)}\bar q(s_g,\zeta _x)=T_{(g,x)}\bar
q(s_{g^{\prime}}^{\prime},\zeta _{x^{\prime}}^{\prime})$ implies that there
exists $\phi\in\frak D_0$ and $\tilde\zeta _{\phi}\in T_{\phi}\frak D_0$
such that
$$(s_{g^{\prime}}^{\prime},\zeta _{x^{\prime}}^{\prime})=(T_gR_{\phi}^{
\frak M}(s_g)+Z_{c_{\phi}(\tilde\zeta _{\phi})}^{\frak M},T_x\phi ^{-1}
(\zeta _x )-c_{\phi}(\tilde\zeta _{\phi})_{\phi ^{-1}(x)})\tag3.30$$
and thus
$$\bar P_{[g,x]}(T_{(g,x)}\bar q(s_g,\zeta _x)):=T_{(g,x)}\bar q\circ
P_{(g,x)}(s_g,\zeta _x)\tag3.31$$
gives a well defined connection on $\frak M\times _{\frak D_0}M\rightarrow
\frak M/\frak D_0$.\qed\enddemo
There is, however, another connection on $\frak M\times TM$ which is induced
from $\omega$. In fact, let $\tilde U$
denote the horizontal lift of $U\in\frak X(\frak M\times M)$ with respect to
$\omega$. Then the induced covariant derivative is given by the formula
$$(\nabla _UV)(g,x)=(0_g,[u,(i_{\tilde U}\ d_{\frak M\times LM^+}\ \hat V)
(g,u)]).\tag3.32$$
\proclaim{Proposition 10} Let $P$ be the connection defined in (3.27). Then
the
covariant derivatives $\nabla ^{\text{ver}}$ and $\nabla$ coincide.
\endproclaim
\demo{Proof} Any $U\in\frak X(\frak M\times M)$ can be split into the
following summands
$$U_{(g,x)}=(s_{(g,x)},\zeta _{(g,x)})=(s_{(g,x)}^h,0_x)+(L_{Y^{
\beta _g(s_{(g,x)})}}\hat g,
-\beta _g(s_{(g,x)}))+P_{(g,x)}U_{(g,x)}.\tag3.33$$
If $U\in\Gamma (\frak M\times TM)$ then it is obvious from (3.28) and (3.32)
that $\nabla ^{\text{ver}}$ and $\nabla$ reduce to the Levi Civita connection
on the fibres for the metric $\nu$.\par
For $U_{(g,x)}=(s_{(g,x)},0_x)$, where $s_{(g,x)}\in T_g\frak M$ is
horizontal $\forall x$, and for $V\in\Gamma (\frak M\times TM)$
the commutator reads
$$[U,V]_{(g,x)}=(-\frac{d}{dt}\vert _{t=0}\ s_{(g,Fl_t^{\bar V}(x))},
\frac{d}{dt}\vert _{t=0}\ \bar V_{(g+ts_{(g,x)},x)})\tag3.34$$
where $Fl_t^{\bar V}$ denotes the flow generated by $\bar V_{(g,.)}\in
\frak X(M)$ for fixed $g$. Inserting of (3.34) into
(3.28) gives
$$\split & 2\nu (\nabla _U^{\text{ver}}V,W)_{(g,x)}=s_{(g,x)}(\bar V_{(g,x)},
\bar W_{
(g,x)})+2g_x(\frac{d}{dt}\vert _{t=0}\ \bar V_{(g+ts_{(g,x)},x)},\bar
W_{(g,x)})\\
&=(\hat s_{(g,x)})_{ij}(u)\hat V(g,u)^i\hat W(g,u)^j+2\hat g_{ij}(u)
\frac{d}{dt}\vert _{t=0}\hat V(g+ts_{(g,x)},u)^i\hat W(g,u)^j.\endsplit
\tag3.35$$
The horizontal lift of $U$ is $\tilde U_{(g,u)}=(s_{(g,x)},
Z_{\xi}(u))$, where $\xi _{\ k}^i(u)=
-\frac{1}{2}\hat g^{ij}(u)(\hat s_{(g,x)})_{jk}(u)$ and $\pi (u)=x$.
We derive from (3.32)
$$(\nabla _UV)(g,x)=(0_g,[u,\frac{d}{dt}\vert _{t=0}\hat V(g+ts_{(g,
x)},u)+(L_{Z_{\xi}}\hat V(g,.))(u)]).\tag3.36$$
Since $(L_{Z_{\xi}}\hat V(g,.))^i(u)=\frac{1}{2}\hat g^{ij}(u)(\hat s_{(g,
x)})_{jk}(u)\hat V^k(g,u)$, where the components are taken according to
(2.6), the scalar product of (3.36) and $W$ finally gives (3.35).\par
For $U_{(g,x)}=Z_{\beta _g(s_{(g,x)})}^{\frak M\times M}(g,x)$, where
$Z^{\frak M\times M}$ is the fundamental vector field on $\frak M\times M$,
it is not difficult to see that $[U,V]$ is vertical. So (3.28) implies
$$\split 2\nu (\nabla _U^{\text{ver}}V,W)_{(g,x)} &=\nu _{(g,x)}([Z_{
\beta _g(s_{(g,x)})}^{\frak M\times M},V]_{(g,x)},\bar W_{(g,x)})-
\nu _{(g,x)}([
Z_{\beta _g(s_{(g,x)})}^{\frak M\times M},W]_{(g,x)},\bar V_{(g,x)})\\ &
\qquad +
L_{Z_{\beta _g(s_{(g,x)})}^{\frak M\times M}}(\nu (\bar V,\bar W))(g,x)\\ &=
2g_x((L_{Z_{\beta _g(s_{(g,x)})}^{\frak M\times M}}\bar V)_{(g,x)},\bar
W_{(g,x)}),\endsplit\tag3.37$$
where we have used the fact that $Z_{\zeta}^{\frak M\times M}$ is a Killing
vector field of $\nu$ for arbitrary $\zeta\in\frak X(M)$. \par
On the other hand, using the identity
$$(\widehat{L_{Z_{\zeta}^{\frak M\times M}}\bar V})(g,u)=(L_{Z_{\zeta}
^{\frak M\times LM^+}}\hat V)(g,u)\tag3.38$$
and that the horizontal lift of $U$ is
$\tilde U_{(g,u)}=Z_{\beta _g(s_{(g,x)})}^{\frak M\times LM^+}$, where
$Z^{\frak M\times LM^+}$ is the fundamental vector field on $\frak M\times
LM^+$, one can derive from (3.32)
$$(\nabla _UV)(g,x)=(0_g,[u,(L_{Z_{\beta _g(s_{(g,x)})}^{\frak M\times LM^+}}
\hat V)(g,u)])=(0_g,(L_{Z_{\beta _g(s_{(g,x)})}^{\frak M\times M}}\bar V)_{(
g,x)}).\tag3.39$$
Taking the scalar product with $W$ finally gives the result.
In summary we can conclude $\nabla ^{\text{ver}}=\nabla$.\qed\enddemo
The metric $\nu ^{\prime}$ projects to a metric $\bar\nu ^{\prime}$ on
$\frak M\times _{\frak D_0}M$, which, in fact, is given by the combination
of the metric on $\frak M/\frak D_0$ induced by $<,>$ together with
$\bar\nu$. Let $\nabla ^{\bar\nu ^{\prime}}$ denote its Levi Civita
connection and define $\bar\nabla ^{\text{ver}}:=\bar P\circ\nabla ^{\bar
\nu ^{\prime}}$. Together with proposition 10 this leads to
\proclaim{Corollary 11} Let $\bar\nabla$ be the covariant derivative on
$\frak M\times _{\frak D_0}TM$ induced by the connection $\bar\omega$ on
$\frak M\times _{\frak D_0}LM^+\rightarrow\frak M\times _{\frak D_0}M$.
Then
$\bar\nabla$ and $\bar\nabla ^{\text{ver}}$ coincide.\endproclaim
Let us now display our geometrical data:
\roster
\item The fibre bundle $\frak M\times _{\frak D_0}M\rightarrow\frak M/
\frak D_0$ whose tangent bundle along its fibres is $\frak M\times _{
\frak D_0}TM\rightarrow\frak M\times _{\frak D_0}M$ with curvature $\Omega _
{\bar\omega}$.
\item The vertical metric $\bar\nu$ on $\frak M\times _{\frak D_0}TM$.
\item The connection $\bar P$ on $\frak M\times _{\frak D_0}M\rightarrow
\frak M/\frak D_0$ which induces $\bar\nabla$ on $\frak M\times _{\frak D_0}
TM$.
\item Additionally, one may consider a representation $\tau$ of $Spin(n)$
in order to construct vector
bundles associated to $\frak M\times _{\frak D_0}LM^+$ to which the Dirac
operators
(3.25) couple. Examples are the Rarita Schwinger operator and the signature
operator which corresponds to chiral spin 3/2 fermions and self dual
antisymmetric tensor fields $\text{respectively.}^{{}1,14}$\endroster
The curvature $\Omega ^{(\Cal L)}$ of the determinant line bundle $\Cal L
\rightarrow\frak M/\frak D_0$ associated
with the equivariant family $\lbrace D_{(F,g)}\vert (F,g)\in\Cal G\times
\frak M\rbrace$ can now be calculated by means of
the local index $\text{formula}^{{}11}$
$$\Omega ^{(\Cal L)}=\left[2\pi i\int _M\ \hat A(\Omega _{\bar\omega})\ ch(
\tau (\Omega _{\bar\omega}))\right]_{(2)}\in\Omega ^2(\frak M/\frak D_0,
\Bbb R)\tag3.40$$
where the 2-form component of the differential form on $\frak M/\frak D_0$
is taken. Here $\hat A$ and $ch$ are the usual polynomials
$$\hat A(\Omega _{\bar\omega})=\left(det(\frac{\Omega _{\bar\omega}/4\pi}{
sinh(\Omega _{\bar\omega}/4\pi)})\right)^{\frac{1}{2}},\qquad ch(\Omega _{
\bar\omega})=tr(e^{i\tau(\Omega _{\bar\omega})/2\pi}).\tag3.41$$
Let $I^m(GL(n))$ denote the space of $ad$ invariant, symmetric multilinear
real valued functions on $gl(n)$ of degree $m$ where we have choosen
$2m-2=n$. Let $Q\in I^m(GL(n))$ be given in such a way that the integrand
of (3.40) can be written in the form
$$\overline{Q(\Omega _{\bar\omega})}:=\left[\hat A(\Omega _{\bar
\omega})\ ch(\tau (\Omega _{\bar\omega}))\right]_{(2m)}
\in\Omega ^{2m}(\frak M\times _{\frak D_0}M,\Bbb R^n),\tag3.42$$
where the $2m$ form component of the integrand has been taken. The form
$\overline{Q(\Omega _{\bar\omega})}$ is uniquely determined by $Q(\Omega _{
\bar\omega})\in\Omega ^{2m}(\frak M\times _{\frak D_0}LM^+)$ with
$\bar\pi ^{\ast}\overline{Q(\Omega _{\bar\omega})}=Q(\Omega _{\bar\omega})$.
\par
In the next step we want to determine the descent equations for combined
consistent Lorentz- and diffeomorphism anomalies within our formalism. We
choose a fixed connection $B$ on $LM^+\rightarrow M$ and extend it to a
connection on $\Cal G\times\frak M\times LM^+$ where we denote it with the
same symbol. Let $\Omega_B\in\Omega ^{(0,0,2)}$ be its curvature.
Since $Q(\Omega _B)=0$ because of dimensional reasons
the descent equations for consistent gravitational anomalies can be
derived from the transgression $\text{formula}^{{}28}$
$$\split Q(\Omega _{\tilde\omega})=Q(\Omega _{\tilde\omega})-Q(\Omega _B) &=
d_{\Cal G\times\frak M\times LM^+}\ TQ(\tilde
\omega ,B)\\ &=d_{\Cal G\times\frak M
\times LM^+}\ \left(m\int _0^1\ dt\ Q(\tilde\omega -B,
\tilde\Omega _t,\ldots ,\tilde\Omega _t)\right),\endsplit\tag3.43$$
where $TQ(\tilde\omega ,B)\in\Omega ^{2m-1}(\Cal G\times\frak M
\times LM^+)$ is
the secondary characteristic form and $\tilde\Omega _t$ is the curvature of
the interpolating connection $\tilde\omega _t=t\tilde\gamma +(1-t)B$.
For $\alpha$ being a connection with curvature $\Omega$ we define $\Omega _t=
t\Omega +\frac{1}{2}(t^2-t)[\alpha ,\alpha]$.
Explicitly, the components of $\tilde\Omega _t$ read
$$\split &\tilde\Omega _{t\ (F,g,u)}^{\ \ (2,0,0)}(\Cal Y_F^1,\Cal Y_F^2)=
\frac{t^2-t}{2}[\Theta _F,\Theta _F](\Cal Y_F^1,\Cal Y_F^2)(u)
\\ &\tilde\Omega _{t\ (F,g,u)}^{\ \ (0,2,0)}(s_g^1,s_g^2)=
t\Omega _{\omega\ (g,F(u))}^{\ \ (2,0)}(s_g^1,s_g^2)+(t^2-t)\\ &\qquad\qquad
\qquad\qquad \times F^{\ast}[i_{Y^{
\beta _g(s_g^1)}}\Gamma ^{a_g}(g)+\rho _g(s_g^1),i_{Y^{
\beta _g(s_g^2)}}\Gamma ^{a_g}(g)+\rho _g(s_g^2)](u)\\
&\tilde\Omega _{t\ (F,g,u)}^{\ \ (0,0,2)}(X_u^1,X_u^2)=(F^{\ast}R(g)_t+
(\Omega _B)_{(1-t)}+t(1-t)[F^{\ast}\Gamma (g),B])_u(X_u^1,X_u^2)\\
&\tilde\Omega _{t\ (F,g,u)}^{\ \ (1,1,0)}(\Cal Y_F,s_g) =(t^2-t)[\Theta _F(
\Cal Y_F)(u),\omega _{(g,F(u))}^{(1,0)}(s_g)]\\
&\tilde\Omega _{t\ (F,g,u)}^{\ \ (1,0,1)}(\Cal Y_F,X_u)=(t^2-t)[\Theta _F(
\Cal Y_F)(u),(F^{\ast}\Gamma (g)-B)_u(X_u)]\\
&\tilde\Omega _{t\ (F,g,u)}^{\ \ (0,1,1)}(s_g,X_u)=t\Omega _{\tilde\omega\
(F,g,u)}^{\ \ (1,1)}(s_g,X_u)+(t^2-t)[\omega _{(g,F(u))}^{(1,0)}
(s_g),(F^{\ast}\Gamma (g)-B)_u(X_u)].\endsplit\tag3.44$$
Notice that $TQ(\tilde\omega ,B)$ is a basic form, i.e. it descends
to a form $\overline{TQ(\tilde\omega ,B)}$ on $\Cal G\times\frak M
\times M$.
With respect to the grading of $\Omega (\Cal G\times\frak M\times LM^+)$
(3.43) can be written in the form
$$\split (Q(\Omega _{\tilde\omega})-Q(\Omega _B))^{(i,j,2m-i-j)}= & d_{
\Cal G}TQ(\tilde\omega ,B)^{(i-1,j,2m-i-j)}+\hat d_{\frak M}TQ(
\tilde\omega ,B)^{(i,j-1,2m-i-j)}\\ &+\hat d_{LM^+}TQ(\tilde\omega ,B)^{(i,j,
2m-i-j-1)}.\endsplit\tag3.45$$
Now we can apply the transgression $\text{procedure.}^{{}14}$ Let
$\overline{Q(\Omega _{
\tilde\omega})}$, $\overline{Q(\Omega _{B})}$ denote the projections of
$Q(\Omega _{\tilde\omega})$ respectively $Q(\Omega _{B})$ on $\Cal G\times
\frak M\times M$. Using proposition 6 and
(3.40) the transgression of $\Omega ^{(\Cal L)}$ gives
$$\split \pi _{\Cal G\times\frak M}^{\ast}\ \Omega ^{(\Cal L)} &=2\pi i\int_M
\ f^{\ast}\overline{Q(\Omega _{\bar\omega})}=2\pi i\int _M\ \overline{Q(
\Omega _{\tilde\omega})}\\ &=2\pi i\int _M\ [\overline{Q(\Omega _{\tilde
\omega})}-\overline{Q(\Omega _B)}]=d_{\Cal G\times \frak M}\left(2\pi i
\int _M\overline{TQ(\tilde\omega ,B)}\right).\endsplit\tag3.46$$
For fixed $g\in\frak M$ the pullback of (3.46) through $i_g$  yields
$$d_{\frak G}\ \left(2\pi i\int_M\overline{i_g^{\ast}TQ(\tilde\omega,
B)}\right)=0,\tag3.47$$
which can be regarded as the consistency condition for the consistent
anomaly. Actually, the fibre integral in (3.47) projects onto the $(1,n)$
component of $\overline{i_g^{\ast}TQ(\tilde\omega,B)}$ with respect
to the bigrading of $\Omega ^{n+1}(\frak G\times LM^+)$.
Using (3.44), (3.20) and (3.21), we obtain for the combined consistent
gravitational anomaly
$$\split i_g^{\ast} TQ(\tilde\omega ,B) & _{(id_{\Cal G},id_{
\frak D_0},u)}^{(1,2m-2)}
(\xi ,\zeta) =m\int _0^1\ dt(Q(\xi +i_{Y^{\zeta}}\Gamma (g),R(g)_t,
\ldots ,R(g)_t)\\ &+(m-1)(t^2-t)Q(\Gamma (g)-B,[\xi +i_{Y^{\zeta}}\Gamma (g),
\Gamma (g)-B],R(g)_t,\ldots R(g)_t)\\ &+(m-1)tQ(\Gamma (g),i_{Y^{\zeta}}
R(g),R(g)_t,\ldots ,R(g)_t))(u),\endsplit\tag3.48$$
where $(\xi ,\zeta )\in Lie\frak G$. This result agrees in form with previous
calculations done in Refs. 14,29. \par
In summary we have seen that the local index theorem determines the
expression for the consistent gravitational anomaly.
\bigskip\bigskip
{\bf IV. Determination of covariant gravitational anomalies}
\bigskip
In this section we want to derive descent equations for covariant
Lorentz- and diffeomorphism anomalies. In particular, we intend to examine
their geometrical structure.\par
We define the following connection on $\Cal G\times\frak M
\times LM^+$ by
$$\eta _{(F,g,u)}(\Cal Y_F,s_g,X_u)=(F^{\ast}\Gamma (g))_u(X_u).
\tag4.1$$
This connection is $\frak G$ invariant but it does not induce a connection on
the universal bundle.
\proclaim{Proposition 12} The curvature $\Omega _{\eta}$ of $\eta$ has the
following components:
$$\split &\Omega _{\eta}^{\ (2,0,0)}=\Omega _{\eta}^{\ (0,2,0)}=
\Omega _{\eta}^{\ (1,1,0)}=0\\ &\Omega _{\eta\ (F,g,u)}^
{\ \ (0,0,2)}(X_u^1,X_u^2)=(F^{\ast}
R(g))_u(X_u^1,X_u^2)\\ &\Omega _{\eta\ (F,g,u)}^{\ \ (1,0,1)}(
\Cal Y_F,X_u)=(d_{F^{\ast}\Gamma (g)}\Theta _F(\Cal Y_F))_u(X_u)\\ &
\Omega _{\eta\ (F,g,u)}^{\ \ (0,1,1)}
(s_g,X_u)=(F^{\ast}\Xi _g(s_g))_u(X_u).\endsplit\tag4.2$$\endproclaim
\demo{Proof} Only the (1,0,1) component of $\Omega _{\eta}$ remains to be
calculated. Using (3.21) this follows from
$$\split (d_{\Cal G}\ \eta _{(.,g,u)})_F(\Cal Y_F)(X_u) &=
\frac{d}{dt}\vert _{t=0}((exp\ t\Theta _F(\Cal Y_F))^{\ast}F^{\ast}
\Gamma (g))_u(X_u)\\ &=(L_{Z_{\Theta _F(\Cal Y_F)}}F^{\ast}
\Gamma (g))_u(X_u)=(d_{F^{\ast}\Gamma (g)}\Theta _F(\Cal Y_F))_u(X_u).
\endsplit\tag4.3$$\qed\enddemo
Let $Q\in I^m(GL^+(n))$, then the covariant descent equations can be obtained
from the transgression formula
$$Q(\Omega _{\tilde\omega})-Q(\Omega _{\eta}) =d_{\Cal G\times\frak M
\times LM^+}\ TQ(\tilde\omega ,\eta ).\tag4.4$$
The covariant anomaly enters the calculation concerning consistent
gravitational anomalies if in the transgression formula (3.46)
$\pi _{\Cal G\times\frak M}^{\ast}\Omega ^{(\Cal L)}$ is replaced with
$\pi _{\Cal G\times\frak M}^{\ast}\Omega ^{(\Cal L)}-\int _M\overline{Q(
\Omega _{\eta})}$.\par
With respect to the triple grading of $\Omega (\Cal G\times\frak M
\times LM^+)$ we obtain from (4.4)
$$\split (Q(\tilde\Omega )-Q(\Omega _{\eta}))^{(i,j,2m-i-j)}= & d_{
\Cal G}
TQ(\tilde\omega ,\eta )^{(i-1,j,2m-i-j)}+\hat d_{\frak M}TQ(
\tilde\omega ,\eta )
^{(i,j-1,2m-i-j)}\\ &+\hat d_{LM^+}\ TQ(\tilde\omega ,\eta )^{(i,j,
2m-i-j-1)}.\endsplit\tag4.5$$
Using (4.3) and
$$\split (d_{\Cal G}\ \tilde\omega _{(.,g,u)}^{(0,1,0)})_F(\Cal Y_F)(s_g) &=
\frac{d}{dt}\vert _{t=0}\ \omega _{(g,Fl_t^{\Cal Y}(F)u)}^{(1,0)}(s_g)
\\
&=-[\Theta _F(\Cal Y_F)(u),(i_{Y^{\beta _g(s_g)}}\Gamma ^{a_g}(g)+\rho _g
(s_g))(u)]\endsplit\tag4.6$$
the components of the curvature of the interpolating connection
$\gamma _t=t\tilde\omega +(1-t)\eta$ are given by
$$\split &\Omega _{t\ (F,g,u)}^{\ \ (2,0,0)}(\Cal Y_F^1,\Cal Y_F^2)=\frac{
t(t-1)}{2}[\Theta ,
\Theta ](\Cal Y_F^1,\Cal Y_F^2)(u)\\ &\Omega _{t\ (F,g,u)}^{\ \ (0,2,0)}(
s_g^1,s_g^2)=\tilde\Omega _{t\ (F,g,u)}^{\ \ (0,2,0)}(s_g^1,s_g^2)\\
&\Omega _{t\ (F,g,u)}^{\ \ (0,0,2)}(X_u^1,X_u^2)=
(F^{\ast}R(g))_u(X_u^1,X_u^2)\\ &\Omega _{t\ (F,g,u)}^{\ \ (1,1,0)}(\Cal Y_F,
s_g )=
t(t-1)[\Theta _F(\Cal Y_F)(u),\omega _{(g,F(u))}^{(1,0)}(s_g)]\\
&\Omega _{t\ (F,g,u)}^{\ \ (1,0,1)}(\Cal Y_F,X_u)=(1-t)(d_{F^{\ast}
\Gamma (g)}\Theta _F(
\Cal Y_F))_u(X_u)\\ &\Omega _{t\ (F,g,u)}^{\ \ (0,1,1)}(s_g,X_u)=F^{\ast}(
\Xi _g(
s_g)-td_{\Gamma (g)}(\omega _{(g,.)}^{(1,0)}(s_g)))_u(X_u).\endsplit
\tag4.7$$
If we restrict (4.4) to a $\frak G$ orbit through $(id_{\Cal G},g)$ and
consider the decomposition according to the bigrading of $\Omega (\frak G
\times LM^+)$ we find the following system of descent equations
$$(Q(i_g^{\ast}\Omega _{\tilde\omega})-Q(i_g^{\ast}\Omega _{\eta}))^{(k,
2m-k)}= d_{\frak G}TQ(i_g^{\ast}\tilde\omega ,i_g^{\ast}\eta )^{(k-1,2m-k)}
+\hat d_{LM^+}TQ(i_g^{\ast}\tilde\omega ,i_g^{\ast}\eta )^{(k,2m-k-1)}.
\tag4.8$$
Explicitly, we find for the non-integrated combined covariant anomaly
$$TQ(i_g^{\ast}\tilde\omega ,i_g^{\ast}\eta )_{(id_{\frak G},u)}^{(1,2m-2)}
=m\ Q(\Theta +i_{Y^c}\Gamma (g),R(g),\ldots ,R(g))(u).\tag4.9$$
For $k=1$ the descent system (4.8) gives
$$0=i_g^{\ast}(Q(\tilde\Omega )-Q(\Omega _{\eta}))^{(1,
n+1)}=\hat d_{LM^+}TQ(i_g^{\ast}\tilde\omega ,i_g^{\ast}\eta )^{(1,n)}.
\tag4.10$$
So the covariant anomaly induces a
class $[\overline{TQ(i_g^{\ast}\tilde\omega ,i_g^{\ast}\eta )}^{(1,n)}]$ in
$H_{d_M}^{(1,n)}(\frak G\times M,\Bbb R)$. This result is similar to that
for the covariant Yang-Mills $\text{anomaly.}^{{}19,30}$ In fact, the form
$TQ(i_g^{
\ast}\tilde\omega ,i_g^{\ast}\eta )^{(1,n)}$ is local in the sense of
Ref. 30,
i.e. it is polynomial in $\Gamma (g)$, $R(g)$ and linear in the ghost
$\Theta +i_{Y^c}\Gamma(g)$. Thus it induces a class in the local
de Rham cohomology of $LM^+$ with one ghost.\par
Additionally, the non-integrated covariant Schwinger term is given by
$$\split TQ(i_g^{\ast}\tilde\omega ,i_g^{\ast}\eta )_{(id_{\frak G},u)}
^{(2,n-1)} &=
m(m-1)\ Q(\Theta +i_{Y^c}\Gamma (g),i_{Y^c}R(g),R(g)\ldots ,R(g))\\ &+
\frac{m(m-1)}
{2}\ Q(\Theta +i_{Y^c}\Gamma (g),d_{\Gamma (g)}(\Theta +i_{Y^c}\Gamma (g)),
R(g),\ldots ,R(g)).\endsplit\tag4.11$$
Notice that the first term on the right hand side of (4.11) can be regarded
as a basic $n$ form on
$LM^+$ and thus it disappears if it is restricted to a $n-1$ dimensional
submanifold of $M$.\par
Evidently, pure covariant diffeomorphism anomalies are obtained by
considering
only forms of the type $TQ(i_g^{\ast}\tilde\omega ,i_g^{\ast}\eta )^{(0,j,
2m-j-1)}$. In order to make contact with previous calculations for $j=1$, we
notice that $L_{Y^{\zeta}}\varphi =0$ implies
$$(i_{Y^{\zeta}}\Gamma (g))\varphi =d_{\Gamma (g)}(i_{Y^{\zeta}}
\varphi ).\tag4.12$$
Let $\hat\zeta\in C(LM^+,T^{1,0})$ be the equivariant function corresponding
to $\zeta\in\frak X(M)$. Its components are given by
$$\hat\zeta ^i(u)=e^i(u^{-1}(\zeta _{\pi _{LM^+}(u)}))=e^i(i_{Y^{\zeta}}
\varphi )(u)=(i_{Y^{\zeta}}\varphi )^i(u).\tag4.13$$
Application of $\lambda$ (2.10) on both sides of (4.12) and using (4.13) give
$$(i_{Y^{\zeta}}\Gamma (g) )_{\ k}^i=\hat\zeta _{\ ;k}^i\tag4.14$$
Hence we obtain
$$TQ(i_g^{\ast}\tilde\omega,i_g^{\ast}\eta )_{(id_{\frak G},u)}^{(0,1,n)}
(\zeta )=m\ Q(d_{\Gamma (g)}\hat\zeta ,R(g),\ldots ,R(g)).\tag4.15$$
This result has been found in Refs. 4,8,14 by other methods.
\par
Let us now briefly comment on covariant Lorentz anomalies. Let us fix a
metric $g$. Consider the map
$\bar i_g\colon\Cal G\times SO_g(M) @>>>\Cal G\times\frak M\times LM^+$,
defined by $\bar i_g(F,u)=(F,g,u)$.
Since $Q(\tilde\Omega )^{(i,0,2m-i)}=0$, if $i\neq 0$, (4.5) reduces to
$$-Q(\bar i_g^{\ast}\Omega _{\eta})^{(i,2m-i)}=d_{\Cal G}TQ(\bar i_g^{\ast}
\tilde\omega ,\bar i_g^{\ast}\eta )
^{(i-1,2m-i)}+\hat d_{SO_g(M)}\ TQ(\bar i_g^{\ast}\tilde\omega ,\bar i_g^{
\ast}\eta )^{(i,2m-i-1)},\tag4.16$$
which gives the set of descent equations for pure covariant Lorentz anomalies
of arbitrary ghost degree. The Levi Civita connection becomes $so(n)$ valued
if restricted to $SO_g(M)$. In fact, $\Gamma (g)$ coincides with
$\Gamma ^a(g)$ on the subbundle $SO_g(M)$ for fixed $g$. Explicitly, the
covariant Lorentz anomaly is given by
$$TQ(\bar i_g^{\ast}\tilde\omega ,\bar i_g^{\ast}\eta )_{(id_{\frak G},u)}
^{(1,2m-2)}(\xi )=m\ Q(\xi ,R_a(g),\ldots ,R_a(g))(u),\tag4.17$$
where $R_a(g)$ is the curvature of $\Gamma ^a$ restricted to $SO_g(M)$.
This result has also be derived by path integral $\text{methods.}^{{}3}$
\par
The statement that pure consistent and covariant Lorentz anomalies can also
be obtained on the same ground as gauge anomalies in Yang-Mills theory is now
easy to understand. In fact, under the assumption from above,
$\Cal G=Aut_0(SO_g(M))$ and (3.2) can be reduced to the commutative diagram
$$\CD Aut_0(SO_g(M))\times SO_g(M)
@>ev_g>> LM^+\\ @VVV @VVV \\ Aut_0(SO_g(M))\times M @>\bar{ev}_g>>
\widetilde{LM}^+\endCD\tag4.18$$
with $ev_g(F,u)=F(u)$ and $\bar{ev}_g(F,x)=g(x)$. \par
The connections
$$\split &\omega _{(F,u)}^{\sharp}(\Cal Y_F,X_u):=(ev_g^{\ast}\Gamma
^a(g))_{(F,g)}(\Cal Y_F,X_u)=\Theta _F(\Cal Y_F)(u)+
(F^{\ast}\Gamma ^a(g))_u(X_u)\\ &\eta _{(F,u)}^{\sharp}(\Cal Y_F,X_u):=(
F^{\ast}\Gamma ^a(g))_u(X_u)\endsplit\tag4.19$$
on $Aut_0(SO_g(M))\times SO_g(M)\rightarrow Aut_0(SO_g(M))\times M$, where
$\Gamma ^a(g)$ is restricted to $SO_g(M)$, can be
used to derive the consistent and covariant descent equations for pure
Lorentz anomalies. So they can be seen as gauge anomalies for $SO_g(M)$.
For the consistent case this viewpoint has been adopted in Ref. 16. \par
Let us now determine the counter terms which relate consistent and covariant
gravitational anomalies. These counter terms for anomalies of arbitrary ghost
degree can be found by considering the $\text{identity}^{{}31}$
$$TQ(\tilde\omega ,\eta )=TQ(\tilde\omega ,B)-TQ(\eta ,
B)+d_{\Cal G\times\frak M\times LM^+}\ S_Q(B,\eta ,\tilde\omega ),\tag4.20$$
where $S_Q(B,\eta ,\tilde\omega )=m(m-1)\int _{t_1+t_2\leq 1}dt_1dt_2
Q(\eta -B,\tilde\omega -B,\Omega _{t_2}^{(t_1)},\ldots ,\Omega _{t_2}
^{(t_1)})$. Here $\Omega _{t_2}^{(t_1)}$ is the curvature of
$B+t_1(\eta-B)+t_2(\tilde\omega -B)$. One should remark that $S_Q$ is a basic
form and thus projects to a form on $\Cal G\times\frak M\times M$. Finally
we identify the counter terms with
$$\split\lambda (\tilde\omega ,\eta ,B)^{(i,j,2m-(i+j+1)}=
&TQ(\eta ,B)^{(i,j,2m-(i+j+1)}-d_{\Cal G}S_Q(B,\eta ,\tilde\omega )^{(i-1,j,
2m-(i+j+1))}\\ &-\hat d_{\frak M}S_Q(B,\eta ,\tilde\omega )^{(i,j-1,
2m-(i+j+1))}\\ &-\hat d_{LM^+}\ S_Q(B,\eta ,\tilde\omega )
^{(i,j,2m-(i+j+2))}.\endsplit\tag4.21$$
To be explicit we calculate the counter term for $k=1$.
The components of the curvature $(\Omega _{\eta})_t$ of
$t\eta +(1-t)B$ are given by
$$\split &(\Omega _{\eta})_t^{\ (2,0,0)}=(\Omega _{\eta})_t^{\ (0,2,0)}=
(\Omega _{\eta})_t^{\ (1,1,0)}=0\\
&(\Omega _{\eta})_{t\ (F,g,u)}^{\ \ (0,0,2)}(X_u^1,X_u^2)=(F^{\ast}
R(g)_t +(\Omega _B)_{(1-t)}+t(1-t)[F^{\ast}\Gamma (g),B])_u(X_u^1,X_u^2)\\
&(\Omega _{\eta})_{t\ (F,g,u)}^{\ \ (1,0,1)}(\Cal Y_F,X_u)=t(d_{F^{\ast}
\Gamma (g)}\Theta _F(\Cal Y_F))_u(X_u)\\
&(\Omega _{\eta})_{t\ (F,g,u)}^{\ \ (0,1,1)}(s_g,X_u)=t(F^{\ast}\Xi _g(
s_g))_u(X_u).\endsplit\tag4.22$$
So up to an exact form, the counter term for $k=1$ reads
$$\multline TQ(\tilde\eta ,B)^{(1,2m-2)}\\ =m\int _0^1dt\ tQ\ (F^{
\ast}\Gamma (g)-B,F^{\ast}
\Xi _g(s_g)+d_{F^{\ast}\Gamma (g)}\Theta _F(\Cal Y_F),(\Omega _{\eta})_t^{
\ (0,0,2)},\ldots ,(\Omega _{\eta})_t^{\ (0,0,2)}).\endmultline
\tag4.23$$
This term is the gravitational analogue of the Bardeen Zumino counter term
in Yang-Mills $\text{case.}^{{}4}$\par
Before closing this section we want to give another geometrical
interpretation of the covariant diffeomorphism anomaly. This will be an
extension of a study of covariant Yang-Mills anomalies in terms of
presymplectic $\text{geometry.}^{{}32}$ We begin with an
investigation of the properties of covariant gravitational anomalies under
$\frak G$ transformations. For $\vartheta\in\Omega ^1(\Cal G\times\frak M
\times LM^+)$ we define
$$\kappa (\vartheta)_{(F,g,u)}(\xi ,\zeta ,X_u):=\vartheta _{(F,g,u)}(
\Cal Y_{\xi}^{\text{left}}(F),Z_{\zeta}^{\frak M}(g),X_u).\tag4.24$$
\proclaim{Lemma 13} Let $\hat\alpha _{(\phi ,F^{\prime})}(F,g):=(l(\phi
^{-1})\circ F\circ l(\phi )\circ F^{\prime},\phi ^{\ast}g)$ then
$$\multline
\kappa (\tilde\omega )_{(\hat\alpha _{(\phi ,F^{\prime})}(F,g),u)}(\xi ,
\zeta ,X_u))\\=
\kappa (\tilde\omega )_{(F,g,(l(\phi )\circ F^{\prime})u)}((l(\phi )
\circ F^{\prime})^{-1\ast}\xi ,ad(\phi )\zeta ,T_u(l(\phi )
\circ F^{\prime})X_u)\endmultline\tag4.25$$
and the formula is also true for $\eta$.\endproclaim
Let us define the integrated covariant gravitational anomaly of ghost degree
$k$ by
$$\Cal A_{(F,g)}^{(k)}(\xi ,\zeta ):=\int _N\ \left(\overline{i_g^{\ast}TQ(
\tilde\omega ,\eta )}^{(k,2m-k)}\right)_{(F,id_{\frak D_0})}(\Cal Y_{\xi}^{
\text{left}}(F),\zeta ),\tag4.26$$
where $N$ is an appropriate $2m-k$ dimensional submanifold of $M$.
Then it is obvious from the above Lemma that the following formula holds:
$$\Cal A_{(\hat \alpha _{(\phi ,F^{\prime})}(F,g))}^{(k)}(\xi ,\zeta )=
\Cal A_{(F,g)}^{(k)}((l(\phi )\circ F^{\prime})^{-1\
\ast}\xi ,ad(\phi )\zeta ),\tag4.27$$
Let us write $\tilde\Cal A(g,\zeta )=\Cal A_{(id_{\Cal G},g)}^{(1)}(
0,\zeta )$ for the pure covariant diffeomorphism anomaly.
The closed 2-form on $\frak M$
$$\Cal K_g(s_g^1,s_g^2)=\int _M\overline{Q(\Omega _{\eta})}^{(0,2,n)}(s_g^1,
s_g^2)=m(m-1)\int _M\overline{Q(\Xi _g(s_g^1),\Xi _g(s_g^2),R(g),\ldots ,
R(g))}\tag4.28$$
is $\frak D_0$ invariant and
defines a presymplectic structure on $\frak M$. Notice that $\Cal K$
is also exact. This property is just a consequence of the transgression
formula $Q(\Omega _{\eta})=Q(\Omega _{\eta})-Q(\Omega _B)=d_{\Cal G\times
\frak M\times LM^+}TQ(\eta ,B)$ and does not depend on the topology of
$\frak M$. However, $\Cal K$ is the differential of the gravitational
Bardeen Zumino term (4.23) for pure diffeomorphism anomalies.\par
A direct computation, using $d_{\frak M}R(g)=d_{\Gamma (g)}\Xi$, finally
gives
$$i_{Z_{\zeta}^{\frak M}}\ \Cal K =-
d_{\frak M}\ \tilde\Cal A(\zeta ).\tag4.29$$
According to (4.27), $\tilde\Cal A$ transforms equivariantly under the
adjoint action of $\frak D_0$.
Hence the covariant diffeomorphism anomaly is an equivariant momentum
$\text{map}^{{}32}$
for the $\frak D_0$ action on the presymplectic manifold $\frak M$. The form
$\Cal K$ also represents the obstruction for $\tilde\Cal A$ to fullfil a
gravitational analogue of the Wess Zumino consistency condition. This is
the content of the descent system (4.5) for the values $i=0$, $j=1$.
\bigskip\bigskip
{\bf V. Gravitational BRS, anti-BRS algebra and the covariance condition}
\bigskip
In this section we want to derive a covariance condition for the combined
covariant Lorentz- and diffeomorphism anomalies. The first step will be the
construction of a geometrical realization of the gravitational BRS, anti-BRS
$\text{algebra.}^{{}21}$\par
Let us consider the following commutative diagram:
$$\CD \Cal G\times\frak D_0\times\Cal G\times\frak M\times LM^+ @>\psi >>
\frak M\times LM^+\\
@V\hat\pi VV @V\pi VV\\ \Cal G\times\frak D_0\times\Cal G\times\frak M
\times M @>\bar\psi >>\frak M\times M\endCD\tag5.1$$
where $\psi (F_1,\phi ,F_2,g,u):=(g,(F_2\circ F_1\circ l(\phi))(u))$
and $\bar\psi$ is the induced map. Let $\Cal B$ denote the bundle on the left
hand side of (5.1). This bundle is a certain pull-back bundle of the
universal bundle. In
fact, let us consider a right $\frak G\times\frak G$ action on $\Cal B$,
given by
$$(F_1,\phi ,F_2,g,u)\mapsto ((F_1,\phi )\cdot (F^{\prime},\phi ^{\prime}),
\alpha _{(\bar F,\bar\phi )}(F_2,g,u)),\tag5.2$$
where $((F^{\prime},\phi ^{\prime}),(\bar F,\bar\phi ))\in\frak G\times\frak
G$. Then $\Cal B/(\frak G\times\frak G)$ is isomorphic with the universal
bundle.\par
The space of differential forms $\Omega$ on $\Cal G\times
\frak D_0\times\Cal G\times\frak M\times LM^+$ admits a fivefold grading.
We define the following operators
$$\alignat2 &d_{\Cal G}^{(1)}\colon\Omega ^{(i,j,k,l,m)}\rightarrow
\Omega ^{(i+1,j,k,l,m)} &&\\ &\hat d_{\frak D_0}\colon\Omega ^{(i,j,
k,l,m)}\rightarrow\Omega ^{(i,j+1,k,l,m)}, &&\qquad \hat d_{\frak D_0}:=
(-1)^id_{\frak D_0}\\
&\hat d_{\Cal G}^{(2)}\colon \Omega ^{(i,j,k,l,m)}\rightarrow
\Omega ^{(i,j,k+1,l,m)}, &&\qquad \hat d_{\Cal G}^{(2)}=(-1)^{i+j}d_{\Cal G}
^{(2)}\\ &\hat d_{\frak M}\colon\Omega ^{(i,j,k,l,m)}\rightarrow
\Omega ^{(i,j,k,l+1,m)}, &&\qquad \hat d_{\frak M}:=(-1)^{i+j+k}d_{\frak M}\\
&\hat d_{LM^+}\colon \Omega ^{(i,j,k,l,m)}\rightarrow
\Omega ^{(i,j,k,l,m+1)},&&\qquad \hat d_{LM^+}=(-1)^{i+j+k+l}d_{LM^+},
\tag5.3\endalignat$$
where $d_{\Cal G}^{(1)}$ and $d_{\Cal G}^{(2)}$ denote the exterior
derivative on the corresponding copy of $\Cal G$ in the product space.\par
Now we define a connection on $\Cal B$ by $\hat\omega :=\psi ^{\ast}
\omega$. Since
$$\multline T_{(F_1,\phi ,F_2,g,u)}\psi (\Cal Y_{F_1},\zeta _{\phi},
\Cal Y_{F_2},s_g,
X_u)\\ =\left( s_g,T_u(F_2\circ F_1)\lbrace T_ul(\phi )(X_u+Y_u^{c_{\phi}(
\zeta _{\phi})})+
Z_{\Theta _{F_1}(\Cal Y_{F_1})+ad(F_1^{-1})\Theta _{F_2}(\Cal Y_{F_2})}(u)
\rbrace\right)
\endmultline\tag5.4$$
the components of $\hat\omega$ are given by
$$\split &\hat\omega _{(F_1,\phi ,F_2,g,u)}^{(1,0,0,0,0)}(\Cal Y_{F_1})=
(l(\phi )^{\ast}\Theta _{F_1}(\Cal Y_{F_1}))(u)\\
&\hat\omega _{(F_1,\phi ,F_2,g,u)}^{(0,1,0,0,0)}(\zeta _{\phi})=((F_2\circ
F_1\circ l(\phi ))^{\ast}\Gamma (g))_u(Y_u^{c_{\phi}(\zeta _{\phi})})\\
&\hat\omega _{(F_1,\phi ,F_2,g,u)}^{(0,0,1,0,0)}(\Cal Y_{F_2})=
\Theta _{F_2}(\Cal Y_{F_2})((F_1\circ l(\phi ))u)\\
&\hat\omega _{(F_1,\phi ,F_2,g,u)}^{(0,0,0,1,0)}(s_g)=
(i_{Y^{\beta _g(s_g)}}\Gamma ^{a_g}(g)+\rho _g(s_g))
((F_2\circ F_1\circ l(\phi ))u)\\
&\hat\omega _{(F_1,\phi ,F_2,g,u)}^{(0,0,0,0,1)}(X_u)=((F_2\circ
F_1\circ l(\phi ))^{\ast}\Gamma (g))_u(X_u).\endsplit\tag5.5$$
Using (5.4) and (2.62) the components of the curvature $\hat\Omega$
of $\hat\omega$ read
$$\split &\hat\Omega ^{(2,0,0,0,0)}=\hat\Omega ^{(0,0,2,0,0)}=\hat
\Omega ^{(1,1,0,0,0)}=\hat\Omega ^{(1,0,1,0,0)}=\hat\Omega ^{(1,0,0,1,0)}=
\hat\Omega ^{(1,0,0,0,1)}\\ &=\hat\Omega ^{(0,1,1,0,0)}=\hat\Omega
^{(0,0,1,1,0)}=\hat\Omega ^{(0,0,1,0,1)}=0\\
&\hat\Omega _{(F_1,\phi ,F_2,g,u)}^{(0,2,0,0,0)}(\zeta _{\phi}^1,
\zeta _{\phi}^2)=
((F_2\circ F_1\circ l(\phi ))^{\ast}R(g))_u(Y_u^{c_{\phi}(\zeta _{\phi}^1)},
Y_u^{c_{\phi}(\zeta _{\phi}^2)})\\
&\hat\Omega _{(F_1,\phi ,F_2,g,u)}^{(0,0,0,2,0)}(s_g^1,s_g^2)=
\Omega _{\omega\ (g,(F_2\circ F_1\circ l(\phi ))(u))}^{\ \ (2,0)}(s_g^1,
s_g^2)\\
&\hat\Omega _{(F_1,\phi ,F_2,g,u)}^{(0,0,0,0,2)}(X_u^1,X_u^2)=
((F_2\circ F_1\circ l(\phi ))^{\ast}R(g))_u(X_u^1,X_u^2)\\
&\hat\Omega _{(F_1,\phi ,F_2,g,u)}^{(0,1,0,1,0)}(\zeta _{\phi},s_g)=
((F_2\circ F_1\circ l(\phi ))^{\ast}(\Xi_g^{a_g}(s_g)-d_{\Gamma (g)}(i_{Y^{
\beta _g(s_g)}}\Gamma ^{a_g}(g)))_u(Y_u^{c_{\phi}(\zeta _{\phi})})\\
&\hat\Omega _{(F_1,\phi ,F_2,g,u)}^{(0,0,0,1,1)}(s_g,X_u)=
((F_2\circ F_1\circ l(\phi ))^{\ast}(\Xi_g^{a_g}(s_g)-d_{\Gamma (g)}(i_{Y^{
\beta _g(s_g)}}\Gamma ^{a_g}(g)))_u(X_u)\\
&\hat\Omega _{(F_1,\phi ,F_2,g,u)}^{(0,1,0,0,1)}(\zeta _{\phi},X_u)=
((F_2\circ F_1\circ l(\phi ))^{\ast}R(g))_u(Y_u^{c_{\phi}(\zeta _{\phi})},
X_u).\endsplit\tag5.6$$
Now we can derive the BRS, anti-BRS relations for Lorentz
transformations and diffeomorphisms.
Let $j_g\colon\Cal G\times\frak D_0\times\Cal G
\times\frak D_0
\times LM^+\hookrightarrow\Cal G\times\frak D_0\times\Cal G\times\frak M
\times LM^+$, defined by $j_g(F_1,\phi _1,F_2,\phi _2,u)=(F_1,\phi _1,F_2,
\phi _2^{\ast}g,u)$
denote the embedding through $g\in\frak M$. We shall use the following
abbreviations
$$\alignat2 & (j_g^{\ast}\hat\omega )^{(1,0,0,0,0)} =\Theta &&\qquad
(j_g^{\ast}\hat\Omega )^{(0,0,0,2,0)} =i_{\bar c}i_{\bar c}R\\
&(j_g^{\ast}\hat\omega )^{(0,0,1,0,0)} =\bar\Theta &&\qquad
(j_g^{\ast}\hat\Omega )^{(0,0,0,0,2)}=R\\
& (j_g^{\ast}\hat\omega )^{(0,0,0,1,0)} =i_{\bar c}\Gamma &&\qquad
(j_g^{\ast}\hat\Omega )^{(0,2,0,0,0)} =i_{c}i_{c}R\\
&(j_g^{\ast}\hat\omega )^{(0,1,0,0,0)} =i_{c}\Gamma  &&\qquad
(j_g^{\ast}\hat\Omega )^{(0,1,0,1,0)}=i_{\bar c}i_cR\\
&\delta :=d_{\Cal G}^{(1)}+\hat d_{\frak D_0}^{(1)} &&\qquad
\bar\delta :=d_{\Cal G}^{(2)}+\hat d_{\frak D_0}^{(2)},\tag5.7\endalignat$$
where $d_{\frak D_0}^{(1)}$ and $d_{\frak D_0}^{(2)}$ are the exterior
derivatives
on $\frak D_0$ with respect to the first and second copy of $\frak D_0$
respectively. By considering $j_g^{\ast}\hat\Omega$ the following relations
can be obtained.
$$\split &d_{\Cal G}^{(2)}\Theta =0,\qquad d_{\Cal G}^{(1)}\bar\Theta +
[\Theta ,\bar\Theta ]=0\\
&\delta (\Theta +i_c\Gamma )=-\frac{1}{2}[\Theta +i_c\Gamma ,
\Theta +i_c\Gamma]+i_ci_cR\\ &\bar\delta (\bar\Theta +i_{\bar c}\Gamma )=
-\frac{1}{2}[\bar\Theta +i_{\bar c}\Gamma ,\bar\Theta +i_{\bar c}\Gamma ]+
i_{\bar c}i_{\bar c}R\\ &\delta (\bar\Theta +i_{\bar c}\Gamma )+\bar\delta
(\Theta +i_{c}\Gamma )=-[\Theta +i_{c}\Gamma ,\bar\Theta +i_{\bar c}\Gamma ]
+i_{\bar c}i_cR,\endsplit\tag5.8$$
The equations in the first line of (5.8) reflect the BRS, anti-BRS structure
in Yang-Mills $\text{theory.}^{{}19}$ However, if the combined
transformations (Lorentz
and diffeomorphism) are considered the derivatives $\delta$ and $\bar
\delta$ act symmetrically. Thus we have seen that $\Cal B$ provides an
appropriate geometrical framework to realize the gravitational BRS, anti-BRS
multiplet.
\par
In this extended framework the calculation of consistent anomalies can be
carried out if in the relevant formulas of Sec. III, $\tilde\omega$ is
replaced with $\hat\omega$.\par
In order to study the covariant case we define another connection on
$\Cal B$ by
$$\split &\hat\eta ^{(0,0,1,0,0)}=\hat\eta ^{(0,0,0,1,0)}=0\\
&\hat\eta _{(F_1,\phi ,F_2,g,u)}^{(1,0,0,0,0)}(\Cal Y_{F_1})=(l(\phi )^{\ast}
\Theta _{F_1}(\Cal Y_{F_1}))(u)\\ &\hat\eta _{(F_1,\phi ,F_2,g,u)}
^{(0,1,0,0,0)}(\zeta _{\phi})=((F_2\circ F_1\circ l(\phi ))^{\ast}
\Gamma (g))_u(Y_u^{c_{\phi}(\zeta _{\phi})})\\
&\hat\eta _{(F_1,\phi ,F_2,g,u)}^{(0,0,0,0,1)}(X_u)=((F_2\circ F_1
\circ l(\phi ))^{\ast}\Gamma (g))_u(X_u).\endsplit\tag5.9$$
The components of the corresponding curvature can now been easily calculated.
They read
$$\split &\Omega _{\hat\eta}^{\ (2,0,0,0,0)}=\Omega _{\hat\eta}
^{\ (0,0,2,0,0)}=\Omega _{\hat\eta}^{\ (0,0,0,2,0)}=\Omega _{\hat\eta}
^{\ (1,0,1,0,0)}=\Omega _{\hat\eta}^{\ (1,0,0,1,0)}\\ &=
\Omega _{\hat\eta}^{\ (1,1,0,0,0)}=\Omega _{\hat\eta}^{\ (1,0,0,0,1)}=
\Omega _{\hat\eta}^{\ (0,0,1,1,0)}=0\\
&\Omega _{\hat\eta\ (F_1,\phi ,F_2,g,u)}^{\ \ (0,2,0,0,0)}(\zeta _{\phi}^1,
\zeta _{\phi}^2)=((F_2\circ F_1\circ l(\phi ))^{\ast}R(g))_u(Y_u^{c_{\phi}
(\zeta _{\phi}^1}),Y_u^{c_{\phi}(\zeta _{\phi}^2)})\\
&\Omega _{\hat\eta\ (F_1,\phi ,F_2,g,u)}^{\ \ (0,0,0,0,2)}(X_u^1,X_u^2)=
((F_2\circ F_1\circ l(\phi ))^{\ast}R(g))_u(X_u^1,X_u^2)\\
&\Omega _{\hat\eta\ (F_1,\phi ,F_2,g,u)}^{\ \ (0,1,1,0,0)}(\zeta _{\phi},
\Cal Y_{F_2})=((F_1\circ l(\phi ))^{\ast}\ d_{F_2^{\ast}\Gamma (g)}\
\Theta _{F_2}(\Cal Y_{F_2}))_u(Y_u^{c_{\phi}(\zeta _{\phi})})\\
&\Omega _{\hat\eta\ (F_1,\phi ,F_2,g,u)}^{\ \ (0,0,1,0,1)}(\Cal Y_{F_2},
X_u)=((F_1\circ l(\phi ))^{\ast}\ d_{F_2^{\ast}\Gamma (g)}\
\Theta _{F_2}(\Cal Y_{F_2}))_u(X_u)\\
&\Omega _{\hat\eta\ (F_1,\phi ,F_2,g,u)}^{\ \ (0,1,0,1,0)}(\zeta _{\phi},s_g)
=((F_2\circ F_1\circ l(\phi ))^{\ast}\ \Xi_g(s_g))_u(Y_u^{c_{\phi}(
\zeta _{\phi})})\\
&\Omega _{\hat\eta\ (F_1,\phi ,F_2,g,u)}^{\ \ (0,0,0,1,1)}(s_g,X_u)
=((F_2\circ F_1\circ l(\phi ))^{\ast}\ \Xi_g(s_g))_u(X_u)\\
&\Omega _{\hat\eta\ (F_1,\phi ,F_2,g,u)}^{\ \ (0,1,0,0,1)}(\zeta _{\phi},X_u)
=((F_2\circ F_1\circ l(\phi ))^{\ast}R(g))_u(Y_u^{c_{\phi}(\zeta _{\phi})},
X_u).\endsplit\tag5.10$$
We proceed like in Sec. IV. Let $\gamma _t=t\hat\omega +(1-t)\hat\eta$ be
the interpolating connection then its curvature admits the following
components
$$\split &\Omega _{t}^{\prime (2,0,0,0,0)}=\Omega _{t}^{\prime (1,0,1,0,0)}=
\Omega _{t}^{\prime (1,0,0,1,0)}=\Omega _{t}^{\prime (1,0,0,0,1)}=\Omega _{t}
^{\prime (1,1,0,0,0)}=0\\ &\Omega _{t\ (F_1,\phi ,F_2,g,u)}
^{\prime\ (0,0,2,0,0)}(\Cal Y_{F_1}^1,\Cal Y_{F_2}^2)=\frac{t(t-1)}{2}
(F_1\circ l(\phi ))^{\ast}([\Theta ,\Theta ](\Cal Y_{F_2}^1,\Cal Y_{F_2}^2))
(u)\\ &\Omega _{t\ (F_1,\phi ,F_2,g,u)}^{\ \ (0,0,0,2,0)}(s_g^1,s_g^2)=
\tilde\Omega _{t\ (F_2\circ F_1,g,l(\phi )u)}^{\ \ (0,2,0)}(s_g^,s_g^2)\\
&\Omega _{t\ (F_1,\phi ,F_2,g,u)}^{\prime \ (0,2,0,0,0)}(\zeta _{\phi}^1,
\zeta _{\phi}^2)=((F_2\circ F_1\circ l(\phi ))^{\ast}R(g))_u(\Cal Y_u
^{c_{\phi}(\zeta _{\phi}^1)},\Cal Y_u^{c_{\phi}(\zeta _{\phi}^2)})\\
&\Omega _{t\ (F_1,\phi ,F_2,g,u)}^{\prime\ (0,0,0,0,2)}(X_u^1,X_u^2)=
((F_2\circ F_1\circ l(\phi ))^{\ast}R(g))_u(X_u^1,X_u^2)\\
&\Omega _{t\ (F_1,\phi ,F_2,g,u)}^{\prime\ (0,0,1,1,0)}(\Cal Y_{F_2},s_g)=
t(t-1)(F_1\circ l(\phi ))^{\ast}[\Theta _{F_2}(\Cal Y_{F_2}),F_2^{\ast}
\omega _{(g,.)}^{(1,0)}(s_g)](u)\\
&\Omega _{t\ (F_1,\phi ,F_2,g,u)}^{\prime \ (0,1,1,0,0)}(\zeta _{\phi},
\Cal Y_{F_2})=(1-t)(F_1\circ l(\phi ))^{\ast}(d_{F_2^{\ast}\Gamma (g)}
\Theta _{F_2}(\Cal Y_{F_2}))_u(Y_u^{c_{\phi}(\zeta _{\phi})})\\
&\Omega _{t\ (F_1,\phi ,F_2,g,u)}^{\prime \ (0,0,1,0,1)}(\Cal Y_{F_2},X_u)=
(1-t)(F_1\circ l(\phi ))^{\ast}(d_{F_2^{\ast}\Gamma (g)}\Theta _{F_2}
(\Cal Y_{F_2}))_u(X_u)\\
&\Omega _{t\ (F_1,\phi ,F_2,g,u)}^{\prime\ (0,1,0,1,0)}(\zeta _{\phi},s_g)=
(F_2\circ F_1\circ l(\phi ))^{\ast}(\Xi_g(s_g)-td_{\Gamma (g)}(\omega _{(
g,.)}^{(1,0)}(s_g))_u(Y_u^{c_{\phi}(\zeta _{\phi})})\\
&\Omega _{t\ (F_1,\phi, F_2,g,u)}^{\prime\ (0,0,0,1,1)}(s_g,X_u)=
(F_2\circ F_1\circ l(\phi ))^{\ast}(\Xi_g(s_g)-td_{\Gamma (g)}(\omega _{(
g,.)}^{(1,0}(s_g))_u(X_u)\\
&\Omega _{t\ (F_1,\phi ,F_2,g,u)}^{\prime\ (0,1,0,0,1)}(\zeta _{\phi},X_u)=
(F_2\circ F_1\circ l(\phi ))^{\ast}R(g))_u(Y_u^{c_{\phi}(\zeta _{\phi})},
X_u).\endsplit\tag5.11$$
Let us now consider the corresponding transgression formula
$$Q(\hat\Omega )-Q(\Omega _{\hat\eta})=d_{\Cal G\times\frak D_0\times\Cal G
\times\frak M\times LM^+}\ TQ(\hat\omega ,\hat\eta ),\tag5.12$$
which according to the product structure of $\Cal G\times\frak D_0\times
\Cal G\times\frak M\times LM^+$ admits a decomposition into the following
system of descent equations
$$\split (Q(\hat\Omega ) &-Q(\Omega _{\hat\eta}))^{(i,j,k,l,s)}=d_{\Cal G}
^{(1)}
TQ(\hat\omega ,\hat\eta )^{(i-1,j,k,l,s)}+
\hat d_{\frak D_0}TQ(\hat\omega ,\hat\eta )^{(i,j-1,k,l,s)}\\ &+
\hat d_{\Cal G}^{(2)}TQ(\hat\omega ,\hat\eta )
^{(i,j,k-1,l,s)}+\hat d_{\frak M}TQ(\hat\omega ,\hat\eta )^{(i,j,k,
l-1,s)}+\\ &+\hat d_{LM^+}\
TQ(\hat\omega ,\hat\eta )^{(i,j,k,l,s-1)},\endsplit\tag5.13$$
where $s=2m-(i+j+k+l)$. From (5.6), (5.10) and (5.11) it is obvious that
$$\split & Q(\hat\Omega )^{(i,j,k,l,2m-(i+j+k+l)}=Q(\Omega _{\hat\eta})
^{(i,j,k,l,2m-(i+j+k+l)}\\ &=
TQ(\hat\omega ,\hat\eta )^{(i,j,k,l,2m-(i+j+k+l+1)}=0,\qquad\ i\neq 0.
\endsplit\tag5.14$$
A comparison of (5.11) with (4.7) gives
$$TQ(j_g^{\ast}\hat\omega ,j_g^{\ast}\hat\eta )_{(id_{\Cal G},id_{
\frak D_0},F,\phi ,u)}^{(0,0,i,j,2m-i-j-1)}=TQ(i_g^{\ast}\tilde\omega ,
i_g^{\ast}\eta )_{(F,\phi ,u)}^{(i,j,2m-i-j-1)}.\tag5.15$$
Thus we have recovered the usual expression for the covariant anomalies.
Setting $i=1$ we obtain from (5.13)
$$d_{\Cal G}^{(1)}TQ(\hat\omega ,\hat\eta )^{(0,j,k,l,2m-(j+k+l+1)}=0.
\tag5.16$$
If the pull-back by $j_g$ is taken, (5.16) can be viewed as
strong covariance condition for combined Lorentz- and diffeomorphism
anomalies under Lorentz transformations.\par
Using (5.6), (5.10) and (5.11) a lenghty calculation gives
$$\split & i_{Y^{c_{\phi}(\zeta _{\phi})}}TQ(\hat\omega ,\hat\eta )^{(0,0,
k,l,2m-(k+l+1))}=i_{\zeta _{\phi}}TQ(\hat\omega ,\hat\eta )^{(0,1,k,l,
2m-(k+l+2))}\\
& i_{Y^{c_{\phi}(\zeta _{\phi})}}Q(\hat\Omega )^{(0,0,k,l,2m-(k+l))}=
i_{\zeta _{\phi}}Q(\hat\Omega )^{(0,1,k,l,2m-(k+l+1))}\\
& i_{Y^{c_{\phi}(\zeta _{\phi})}}Q(\Omega _{\hat\eta})^{(0,0,k,l,2m-(k+l))}=
i_{\zeta _{\phi}}Q(\Omega _{\hat\eta})^{(0,1,k,l,2m-(k+l+1))}.\endsplit
\tag5.17$$
Here $i_{Y^{c_{\phi}(\zeta _{\phi})}}$ is the substitution operator on
$LM^+$
whereas $i_{\zeta _{\phi}}$ denotes the substitution operator on $\frak D_0$.
If we apply $i_{Y^{c_{\phi}(\zeta _{\phi})}}$ on (5.13) for index values
$i=0$, $j=0$ and compare the result with the result obtained by applying $i_{
\zeta _{\phi}}$ on (5.13) for index values $i=0$, $j=1$ we find
$$i_{\zeta _{\phi}}d_{\frak D_0}TQ(\hat\omega ,\hat\eta )^{(0,0,k,l,
2m-(k+l+1))}=(-1)^{k+l}
L_{Y^{c_{\phi}(\zeta _{\phi})}}TQ(\hat\omega ,\hat\eta )^{(0,0,k,l,2m-
(k+l+1))}.\tag5.18$$
If we combine (5.16) and (5.18) and take the pull-back by $j_g$ we obtain
$$\delta\ TQ(j_g^{\ast}\hat\omega ,j_g^{\ast}\hat\eta )^{(0,0,k,l,2m-(k+
l+1))}=(-1)^{k+l}L_{Y^{c(.)}}TQ(j_g^{\ast}\hat\omega ,j_g^{\ast}\hat\eta )
^{(0,0,k,l,2m-(k+l+1))}.\tag5.19$$
Eq.(5.19) is the strong covariance condition for combined covariant
diffeomorphism- and Lorentz anomalies. For pure diffeomorphism
anomalies this condition agrees with a corresponding one which has previously
been derived in Ref.8. Hence
we have found a geometrical interpretation of this condition in terms of the
geometry of the bundle $\Cal B$.
\bigskip\bigskip
{\bf Acknowledgement}
\bigskip
I would like to thank Prof. J. Wess for the kind hospitality at the LMU
M\"unchen.
\bigskip
\bigskip
\head References\endhead
\bigskip
\roster
\item"{${}^1$}" L. Alvarez-Gaum\'e and E. Witten, Nucl. Phys. {\bf B234},
269 (1984).
\item"{${}^2$}" L.N. Chang and H.T. Nieh, Phys.
Rev. Lett. {\bf 53}, 21 (1984).
\item"{${}^3$}" S. Yajima and T. Kimura, Prog. Theor. Phys. {\bf 77},
1517 (1987).
\item"{${}^4$}" W. Bardeen and B. Zumino, Nucl. Phys. {\bf B244}, 421
(1984).
\item"{${}^5$}" M. Tomiya, Phys. Lett. {\bf 167B}, 411 (1986).
\item"{${}^6$}" M. Ebner, R. Heid and G. Lopes Cardoso, Z. Phys. C-
Particles and Fields {\bf 37}, 85 (1987).
\item"{${}^7$}" M. Abud, J.P. Ader and F. Gieres,
Nucl. Phys. {\bf B339}, 687 (1990).
\item"{${}^8$}" J.P. Ader, F. Gieres and Y. Noirot,
Phys. Lett. {\bf 256B}, 401 (1991).
\item"{${}^9$}" J.M. Bismut, Invent. Math. {\bf 83}, 91 (1986).
\item"{${}^{10}$}" J.M. Bismut and D.S. Freed, Commun.
Math. Phys. {\bf 106}, 159 (1986).
\item"{${}^{11}$}" J.M. Bismut and D.S. Freed, Commun.
Math. Phys. {\bf 107}, 103 (1986).
\item"{${}^{12}$}" M.F. Atiyah and I.M. Singer, Ann. of Math. {\bf 93},
119 (1971).
\item"{${}^{13}$}" M.F. Atiyah and I.M. Singer, Proc. Natl. Acad. Sci. USA
{\bf 81}, 2597 (1984).
\item"{${}^{14}$}" L. Alvarez-Gaum\'e and P. Ginsparg, Ann. Phys. {\bf 161},
423 (1985).
\item"{${}^{15}$}" O. Alvarez, I.M. Singer and B. Zumino, Commun. Math.
Phys. {\bf 96}, 409 (1984).
\item"{${}^{16}$}" L. Bonora, P. Cotta-Ramusino, M. Rinaldi and J. Stasheff,
Commun. Math. Phys. {\bf 112}, 237 (1987).
\item"{${}^{17}$}" D.S. Freed,
Commun. Math. Phys. {\bf 107}, 483 (1986).
\item"{${}^{18}$}" L. Baulieu and I.M. Singer, Commun. Math. Phys. {\bf 135},
253 (1991).
\item"{${}^{19}$}" G. Kelnhofer, to be published in J. Math. Phys. (1993)
\item"{${}^{20}$}" D. Bao and V.P. Nair, Commun. Math. Phys. {\bf 101},
437 (1985).
\item"{${}^{21}$}" L. Baulieu and M. Bellon, Nucl. Phys. {\bf B266},
75 (1986).
\item"{${}^{22}$}" D.G. Ebin, {\it Proc. Symposia in Pure Mathematics\/},
Am. Math. Soc. Vol 15,(Providence, RI AMS, 1970).
\item"{${}^{23}$}" D. Bleecker, {\it Gauge theories and variational
principles\/} (Addison-Wesley, London, 1981).
\item"{${}^{24}$}" S. Kobayashi and K. Nomizu, {\it Foundations of
differential
geometry I\/} (Interscience, New York, 1963).
\item"{${}^{25}$}" T.P. Killingback, Commun. Math. Phys. {\bf 100},
267 (1985).
\item"{${}^{26}$}" L. Baulieu and J. Thierry-Mieg,
Phys. Lett. {\bf 145B}, 53 (1984).
\item"{${}^{27}$}" L. Dabrowski and R. Percacci, Commun. Math. Phys.
{\bf 106}, 691 (1986).
\item"{${}^{28}$}" S.S. Chern, {\it Complex manifolds without potential
theory\/} (Springer, New York, 1979).
\item"{${}^{29}$}" F. Langouche, T. Sch\"ucker and R. Stora,
Phys. Lett. {\bf 145B}, 342 (1984).
\item"{${}^{30}$}" L. Bonora and P. Cotta-Ramusino, Phys. Rev D
{\bf 33}, 3055 (1986).
\item"{${}^{31}$}" R. Bott, {\it Lectures on algebraic and differential
topology\/} (Lect. Notes in Math. 279, Berlin, New York, 1972).
\item"{${}^{32}$}" P. Libermann and C.M. Marle, {\it Symplectic geometry and
analytical mechanics\/} (D. Reidel, Dordrecht, Boston, 1987).
\endroster

\enddocument